\documentclass[fleqn,usenatbib]{mnras}

\usepackage{newtxtext,newtxmath}

\usepackage[T1]{fontenc}
\usepackage{ae,aecompl}

\usepackage{graphicx}	
\usepackage{amsmath}	
\usepackage{amssymb}	

\title[Effects of dynamical dark energy on large-scale structure]{The BAHAMAS project: Effects of dynamical dark energy on large-scale structure}

\author[S. Pfeifer et al.]{
Simon Pfeifer$^{1}$\thanks{E-mail: s.pfeifer@2012.ljmu.ac.uk},
Ian G. McCarthy$^{1}$\thanks{E-mail: i.g.mccarthy@ljmu.ac.uk},
Sam G. Stafford$^{1}$,
Shaun T. Brown$^{1}$,
\newauthor Andreea S. Font$^{1}$,
Juliana Kwan$^{1}$,
Jaime Salcido$^{1}$,
Joop Schaye$^{2}$
\\
$^{1}$Astrophysics Research Institute, Liverpool John Moores University, 146 Brownlow Hill, Liverpool L3 5RF\\
$^{2}$Leiden Observatory, Leiden University, P. O. Box 9513, 2300 RA Leiden, The Netherlands
}

\date{Accepted XXX. Received YYY; in original form ZZZ}

\pubyear{2020}

\begin{document}
\label{firstpage}
\pagerange{\pageref{firstpage}--\pageref{lastpage}}
\maketitle

\begin{abstract}
In this work we consider the impact of spatially-uniform but time-varying dark energy (or `dynamical dark energy', DDE) on large-scale structure in a spatially flat universe, using large cosmological hydrodynamical simulations that form part of the BAHAMAS project.  As DDE changes the expansion history of the universe, it impacts the growth of structure.  We explore variations in DDE that are constrained to be consistent with the cosmic microwave background.  We find that DDE can affect the clustering of matter and haloes at the $\sim10\%$ level (suppressing it for so-called `freezing' models, while enhancing it for `thawing' models), which should be distinguishable with upcoming large-scale structure surveys.  DDE cosmologies can also enhance or suppress the halo mass function (with respect to $\Lambda$CDM) over a wide range of halo masses.  The internal properties of haloes are minimally affected by changes in DDE, however. Finally, we show that the impact of baryons and associated feedback processes is largely independent of the change in cosmology and that these processes can be modelled separately to typically better than a few percent accuracy.
\end{abstract}




\section{Introduction}

The direct detection of the accelerated expansion of the Universe \citep{riess1998,perlmutter1999} ushered in a new era of cosmology and brought with it the standard model of cosmology, the $\Lambda$CDM model, which has been incredibly successful. However, with recent increases in the quantity and quality of observational data, a number of tensions have started to appear that cannot be easily reconciled.  In fact, these tensions have tended to increase in significance with new data and may hint at extra physics that is not encompassed within the standard model of cosmology. 

Perhaps the most well-known tension concerns the expansion rate of space at the present day, $H_0$.  Local measurements of a set of standard candles imply $H_0=74.03\pm1.42$ km $\rm s^{-1}$ $\rm Mpc^{-1}$ \citep{riess2019} and more recently $H_0=73.3\pm1.8$ km km $\rm s^{-1}$ $\rm Mpc^{-1}$ from the measured time delays of gravitationally-lensed quasars \citep{wong2020}, while a combined analysis of cosmic microwave background (CMB) data, baryon acoustic oscillations (BAO) and supernovae have measured $H_0=67.4\pm0.5$ km $\rm s^{-1}$ $\rm Mpc^{-1}$ \citep{planckparams2018}, culminating in a `early vs.~ late-Universe' tension of 5.3$\sigma$ \citep{wong2020}.  Another tension comes from large scale structure (LSS) joint constraints on $\Omega_{\rm m}$ and $\sigma_8$, the mean matter density of the Universe and the linearly-evolved amplitude of matter fluctuations at present day on 8$h^{-1}$Mpc scales, respectively. The {\it Planck} primary CMB data prefers higher values of $\Omega_{\rm m}$ and/or $\sigma_8$ relative to a range of LSS data sets, typically at the 1-3$\sigma$ level (e.g., \citealt{planck_clusters,leauthaud2017,hildebrandt2020}; see \citealt{mccarthy2018} for a recent discussion).

One way of addressing these tensions is through extensions to the $\Lambda$CDM model, which typically add more complex physics and/or relax key assumptions of the model. A popular target is the cosmological constant, $\Lambda$, invoked to explain the observed accelerated expansion of the Universe. Physically-motivated scenarios, such as those based on the scale of particle interactions, suggest a non-zero cosmological constant should be over one hundred orders of magnitude larger than its measured value. Together with the ``coincidence'' problem, i.e. the fact that the energy density of matter and dark energy are of the same order at the current epoch, which requires finely tuned ICs, has led some to argued that the cosmological constant gives a theoretically unsatisfactory explanation for the accelerated expansion of the Universe \citep{weinberg1989}.

The extension focused on in this work is generically termed `dynamical dark energy' (DDE). Instead of modeling dark energy as a cosmological constant, characterised by a constant equation of state parameter with $\textit{w}=-1$, DDE adds an extra degree of freedom by allowing the equation of state parameter to evolve with time; $\textit{w} \longrightarrow \textit{w}(a)$, where $a$ is the expansion factor.  This changes the expansion history of the Universe and subsequently affects the growth of structure.  Therefore, the growth of LSS should serve as an excellent probe of dark energy that is complementary to geometric probes, such as BAO and supernovae, which try to measure the expansion history directly.  In addition, LSS is vital for distinguishing between DDE and modified gravity explanations for the accelerated expansion of the Universe (e.g., \citealt{li2012,mota2018}).

A few methods exist to model LSS statistics. On very large scales one can use linear perturbation theory to calculate the distribution of matter. However, most LSS statistics require accurate modeling on non-linear scales for which this approach is inadequate. A more common approach is to use collisionless simulations to calibrate the so-called ``halo model'' \citep{peacock2000,seljak2000,cooray2002,mead2015}, or to use these simulations to empirically correct linear perturbation models (e.g., \citealt{takahashi2012}). These approaches, which can be accurate to $\approx5\%$, are likely to be insufficient for the next generation of observational surveys like LSST \citep{lsst2012} and Euclid \citep{euclid2013}, which aim to be able to measure statistics, such as the non-linear matter power spectrum, to within percent level accuracy \citep{huterer2002,huterer2005,hearin2012}. Additionally, baryons contribute a significant fraction of the total matter content of the Universe that is not modelled beyond the expansion history in the methods mentioned above. It has been shown that baryonic feedback processes not only affect the spatial distribution of baryons but also induce a back-reaction onto the dark matter distribution that should not be ignored \citep{vandaalen2011,velliscig2014,mummery2017,springel2018,chisari2018,mccarthy2018,vandaalen2020}. Hence, hydrodynamical cosmological N-body simulations are the only method that can model the matter distribution accurately and self-consistently down to highly non-linear scales as well as accurately include the effects of baryons.

Many studies have used collisionless simulations to study the effects of dark energy that differ from the cosmological constant on the dark matter distributions. The first studies explored cosmologies with $w\neq-1$ but still constant with time \citep{ma1999,bode2001,lokas2004} and soon after, a variable equation of state parameter was introduced \citep{klypin2003,linderjenkins2003}. For the interested reader, \citet{baldi2012} reviews different theoretical dark energy models along with relevant studies that utilise cosmological simulations. More recently, dark energy has been studied using collisionless simulations in the context of the halo mass function \citep{francis2009,bhattacharya2011,courtin2011,biswas2019}, non-linear power spectrum \citep{francis2009,casarini2009,alimi2010,heitmann2010}, and has been employed in both semi-analytic \citep{takahashi2012,mead2015,cateneo2019} and emulation \citep{kwan2013,heitmann2014,knabenhans2019,harnois-deraps2019} frameworks. Hydrodynamical simulations have also been used, although to much less extent, specifically to investigate the impact of dark energy on galaxy evolution \citep{penzo2014} and cosmic reionization \citep{maio2006}.

The work presented here uses large cosmological hydrodynamical simulations to study the effects of DDE on LSS for the first time. The large box size of our simulations allows us to study a wide variety of LSS statistics and, by including baryonic effects alongside changes in cosmology, we are able to explore the potential degeneracies that exist between them and whether we can model their combined effect. Our chosen cosmologies are consistent with the latest CMB data and we can therefore ask whether the effect in the LSS statistics between the different cosmologies are distinguishable with current and future LSS surveys. 

This paper is organised as follows: Section~\ref{sec:simulations} presents an overview of the simulations, a brief theoretical background to DDE and explains the parameter selection for our chosen cosmologies. In Section~\ref{sec:lss} we examine LSS clustering statistics, the abundance of haloes and in Section~\ref{sec:halostructure} we show statistics of the internal properties of haloes. We investigate the separability of cosmological and baryonic effects on these statistics in Section \ref{sec:separability}; i.e., we determine to what extent the impact of baryons is dependent upon the choice of cosmology.  Finally, in Section~\ref{sec:discussionresults} we summarise and discuss our results.

\section{Simulations}
\label{sec:simulations}
We use a modified version of the \texttt{BAHAMAS} cosmological hydrodynamical simulation code that includes a prescription of DDE and massive neutrinos. Below we provide a brief overview of the simulations, but the reader should refer to \citet{mccarthy2017} and \citet{mccarthy2018} for a more detailed discussion of the simulations, calibration and comparisons to observations. We describe the theoretical background to the DDE prescription and its implementation in Section~\ref{sec:dde} and the method for choosing suitable cosmological parameters in Section~\ref{sec:cosmoparam}.

\subsection{BAHAMAS}
\label{sec:bahamas}
The simulations were run with the \texttt{BAHAMAS} cosmological  hydrodynamical simulation code and consist of 6 simulations with a periodic box of 400 comoving Mpc/$h$ on a side and containing $2\times1024^3$ particles, equally split between dark matter and baryons. We have also run corresponding collisionless (`dark matter-only') simulations, resulting in a total of 12 simulations.  Initial conditions (ICs) were generated using a modified version of \texttt{N-GenIC}\footnote{\url{https://github.com/sbird/S-GenIC}} \citep{bird2017} with transfer functions at a starting redshift of $z=127$ computed by \texttt{CAMB}\footnote{\url{http://camb.info/}} \citep{lewis1999}. Note that \texttt{CAMB} was compiled with the parameterized post-Friedmann description of cosmic acceleration which allows for a dark energy description that can smoothly cross the phantom divide \citep{hu2007,fang2014}. The same random phases were used to generate each set of ICs allowing for comparisons between the different simulation runs without the complication of cosmic variance.  As in previous BAHAMAS runs, separate transfer functions are used for each constituent, i.e. CDM, baryons and neutrinos, to generate the ICs \citep{bird2020}.

The simulations use a modified version of the Lagrangian TreePM-SPH code \texttt{GADGET3} (last described in \citealt{gadget}), which was modified to include new subgrid physics as part of the \texttt{OWLS} project \citep{schaye2010}. They include an extension for massive neutrinos described in \citet{alihaimound2013} that computes neutrino perturbations on the fly at every time step using a linear perturbation integrator sourced from the non-linear baryons+CDM potential, adding the result to the total gravitational force. Because the neutrino power is calculated at every time step, the dynamical responses of the neutrinos to the baryons+CDM and vice versa are mutually and self-consistently included. We adopt the minimal neutrino mass, $\Sigma M_{\nu}=0.06$ eV, in this work but the reader can refer to \citet{mummery2017} and \citet{mccarthy2018} for the effects of more massive neutrinos.

Additionally, the radiation energy density is included when computing the background expansion rate. This results in a few percent reduction in the amplitude of the present-day linear matter power spectrum relative to simulations that only include the matter and dark energy components in the background expansion rate. The background cosmology was also modified to include DDE as in detail in Section~\ref{sec:dde}.

The simulations include subgrid prescriptions for metal-dependent radiative cooling \citep{wiersma2009}, star formation \citep{schaye2008}, and stellar evolution, mass loss and chemical enrichment \citep{wiersma2009v2} from Type II and Ia supernovae and Asymptotic Giant Branch stars. The simulations also incorporate stellar feedback \citep{vecchia2008} and a prescription for supermassive black hole growth and AGN feedback \citep{booth2009} (which is a modified version of the model originally developed by \citealt{springel2005}). A discussion of the calibration of the feedback will be presented in Section \ref{sec:separability}.

We used a standard friends-of-friends (FoF) algorithm \citep{davis1983} with linking length of 0.2 in units of mean inter-particle separation on the dark matter distribution to identify haloes. The \texttt{SUBFIND} algorithm \citep{springel2001,dolag2009} was used to identify substructures within the FoF groups using a spherical overdensity method and to calculate properties such as R$_{200,\rm crit}$, the radius of a sphere enclosing a mean density of 200 times the critical density, and M$_{200,\rm crit}$, the mass enclosed within.

\subsection{Dynamical dark energy}
\label{sec:dde}
The cosmological constant, $\Lambda$, which is uniform in time and space, gives rise to a repulsive force that counteracts gravity.  DDE modifies this behaviour by positing that dark energy evolves with time while remaining spatially-uniform. Many physical models have been proposed to accomplish this (e.g., \citealt{ratra1988,wetterich1988,brax1999,wetterich2004}). While $\Lambda$ is described by a constant equation of state parameter, $\textit{w}=-1$, a common parameterisation of DDE was introduced by \cite{chevallier2001} and \cite{linder2003},
\begin{equation}\label{equ:deeos}
    \textit{w}(a) = \textit{w}_0 + \textit{w}_{a}(1-a),
\end{equation}
where $a$ is the expansion factor and $w_{0}$ and $w_{a}$ are free parameters. One can recover $\Lambda$ by setting $w_{0}=-1$ and $w_{a}=0$. The benefits of this parameterisation are that one can generate the expansion histories very easily (as we will show below) and that it can mimic the expansion history of many physical DDE models.

Assuming a spatially flat Universe, the expansion history is described by the Friedman equation
\begin{equation}\label{friedman}
    H^2 = \frac{8\pi G}{3}\rho,
\end{equation}
where $H$ is the Hubble parameter, $G$ the gravitational constant and $\rho$ is the sum of the energy densities of the constituents of the Universe, i.e. matter, radiation and DE. The temporal evolution of the energy density is described by a perfect fluid in the form of a differential equation 
\begin{equation}\label{fluidequation}
    \frac{d\rho}{\rho} = -3(\textit{w}+1)\frac{da}{a}.
\end{equation}
The solutions to Equation~\ref{fluidequation} are simple for matter and radiation with $w=0$ and $w=\frac{1}{3}$, respectively. The solution is more complicated for the dark energy equation of state given in Equation~\ref{equ:deeos}, which has an explicit dependence on $a$, and is given by \cite{linder2003} as
\begin{equation}\label{dedensity}
    \rho_{\rm DE} = \rho_{\rm DE,0}a^{-3(1+\textit{w}_{a}+\textit{w}_0)}e^{-3\textit{w}_{a}(1-a)},
\end{equation}
where $\rho_{\rm de,0}$ is the dark energy density at the present day. Substituting Equation~\ref{dedensity} along with the relation for the dimensionless density parameter $\Omega=\frac{8\pi G}{3H_0^2}\rho_0$ for each species into Equation~\ref{friedman} gives an expression for the expansion history as a function of present day energy densities,
\begin{equation}\label{expansionhistory}
    H(a)^2 = H_0^2\left(\Omega_{\rm r}a^{-4} + \Omega_{\rm m}a^{-3} + \Omega_{\rm DE}a^{-3(1+\textit{w}_{a}+\textit{w}_0)}e^{-3\textit{w}_{a}(1-a)}\right).
\end{equation}
Equation~\ref{expansionhistory} was implemented into the BAHAMAS simulations to include the effects of DDE.

\subsection{Cosmological parameter selection}
\label{sec:cosmoparam}
The choice of cosmological parameters is a non-trivial issue and a few factors must be considered during the selection. Cosmological simulations are expensive to run and thus only a relatively small number of different cosmologies can be explored. One option is to pick a fiducial model and simply vary the dark energy parameters over a range of values while keeping the rest of the cosmological parameters fixed. However, this \textit{ad hoc} approach would result in cosmologies that are neither physically-motivated nor consistent with observational constraints. Our approach is to use observational data to constrain the available $w_{0}-w_{a}$ parameter space. The rest of the cosmological parameters (e.g., $H_0$, $\Omega_{\rm m}$, etc.) are chosen to be consistent with observational data by insisting that the cosmological model reproduces our chosen observational data set(s) to within some tolerance.  In this way we can generate cosmologies that are consistent with observations and that allow us to explore a range of DDE behaviours.

\begin{figure}
	\includegraphics[width=\columnwidth]{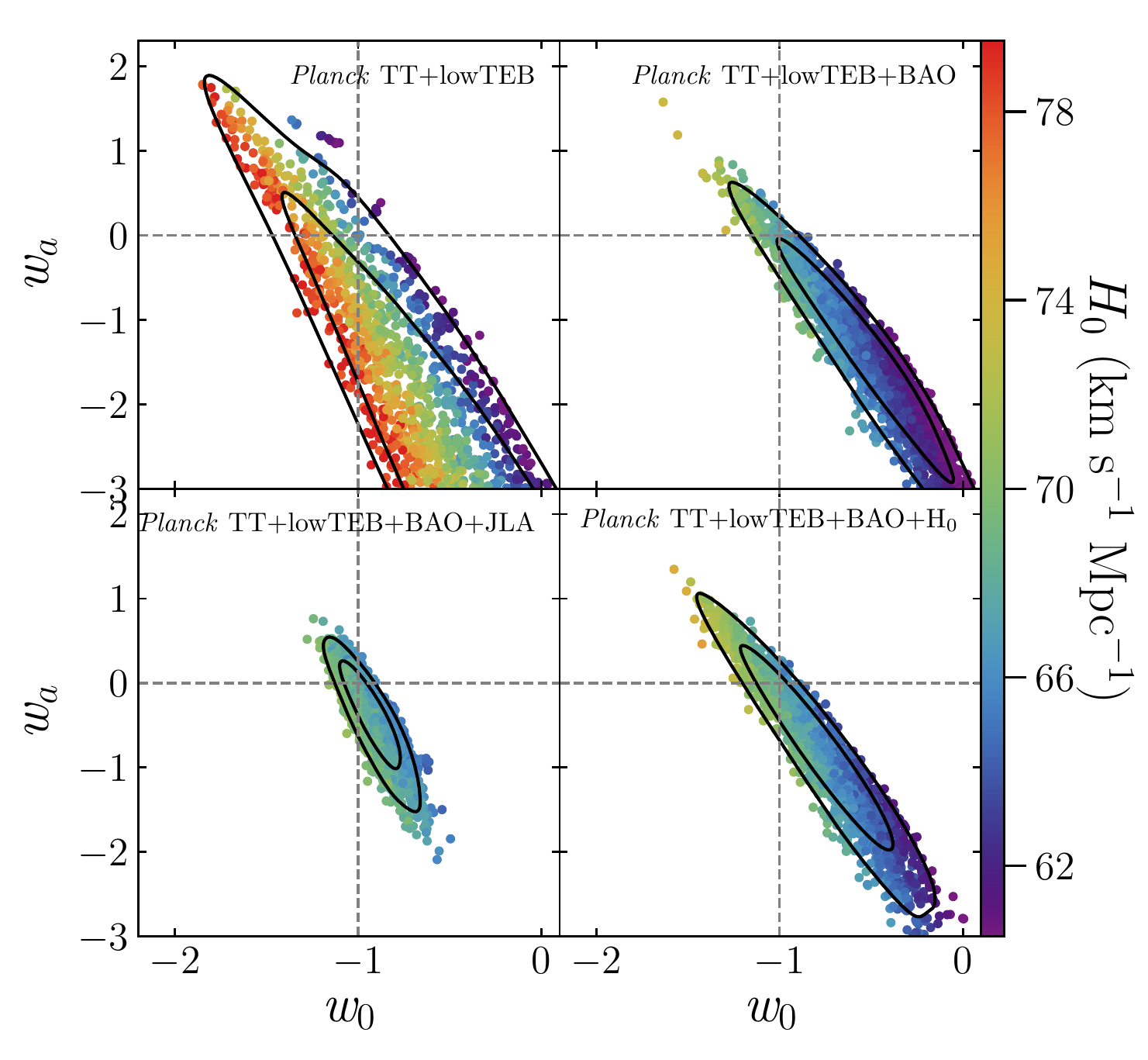}
    \caption{The constraints in the w$_0$-w$_a$ parameter space, in the form of 1$\sigma$ and 2$\sigma$ contours, from different combinations of data. Planck TT+lowTEB (top left) + BAO (top right) + JLA (bottom left)/+ local $H_0$ constraints (bottom right). Points are coloured depending on their $H_0$ value, the dashed lines cross at the cosmological constant and A$_{\rm lens}=1$}
    \label{fig:simruns}
\end{figure}

\begin{figure}
	\includegraphics[width=\columnwidth]{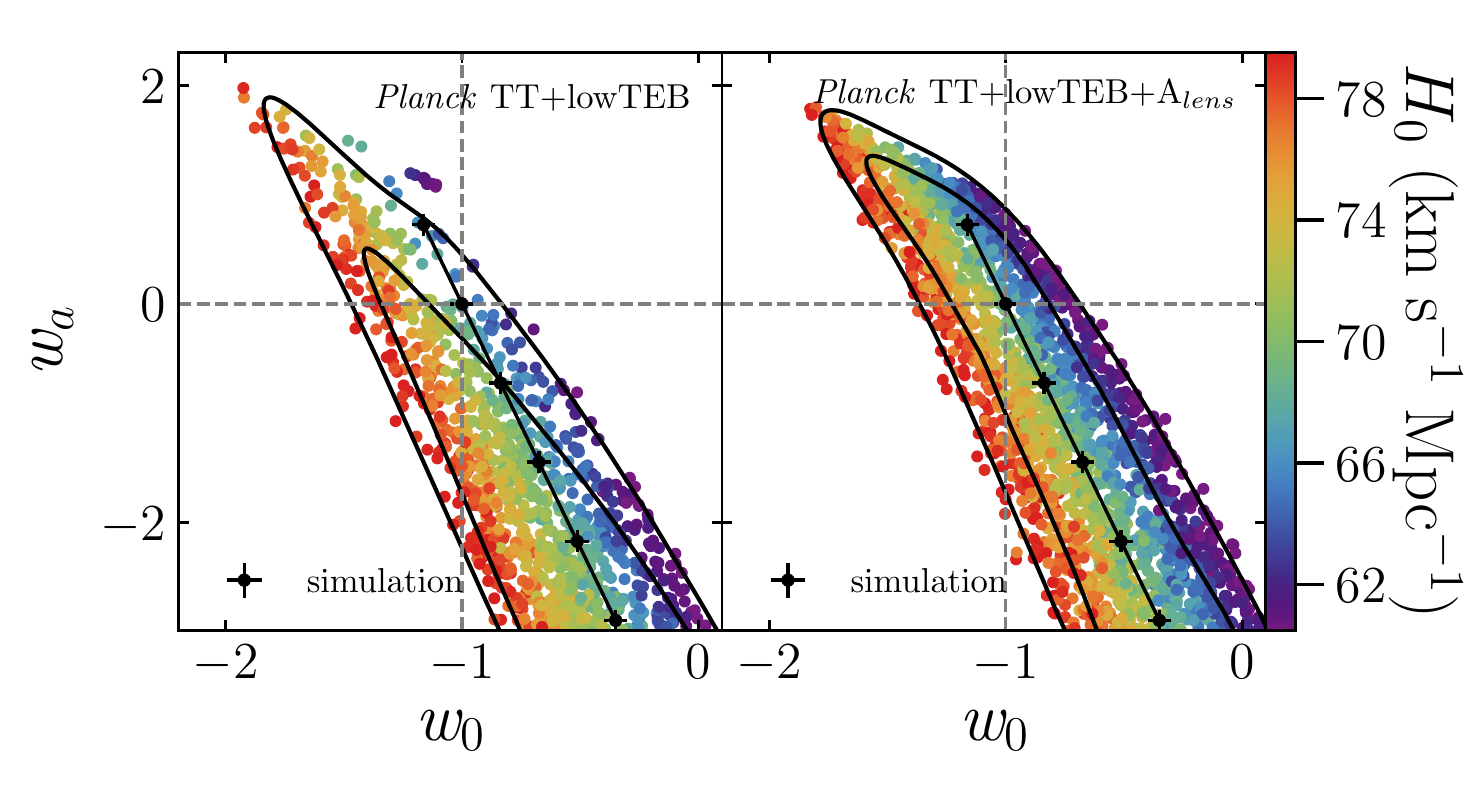}
	\includegraphics[width=\columnwidth]{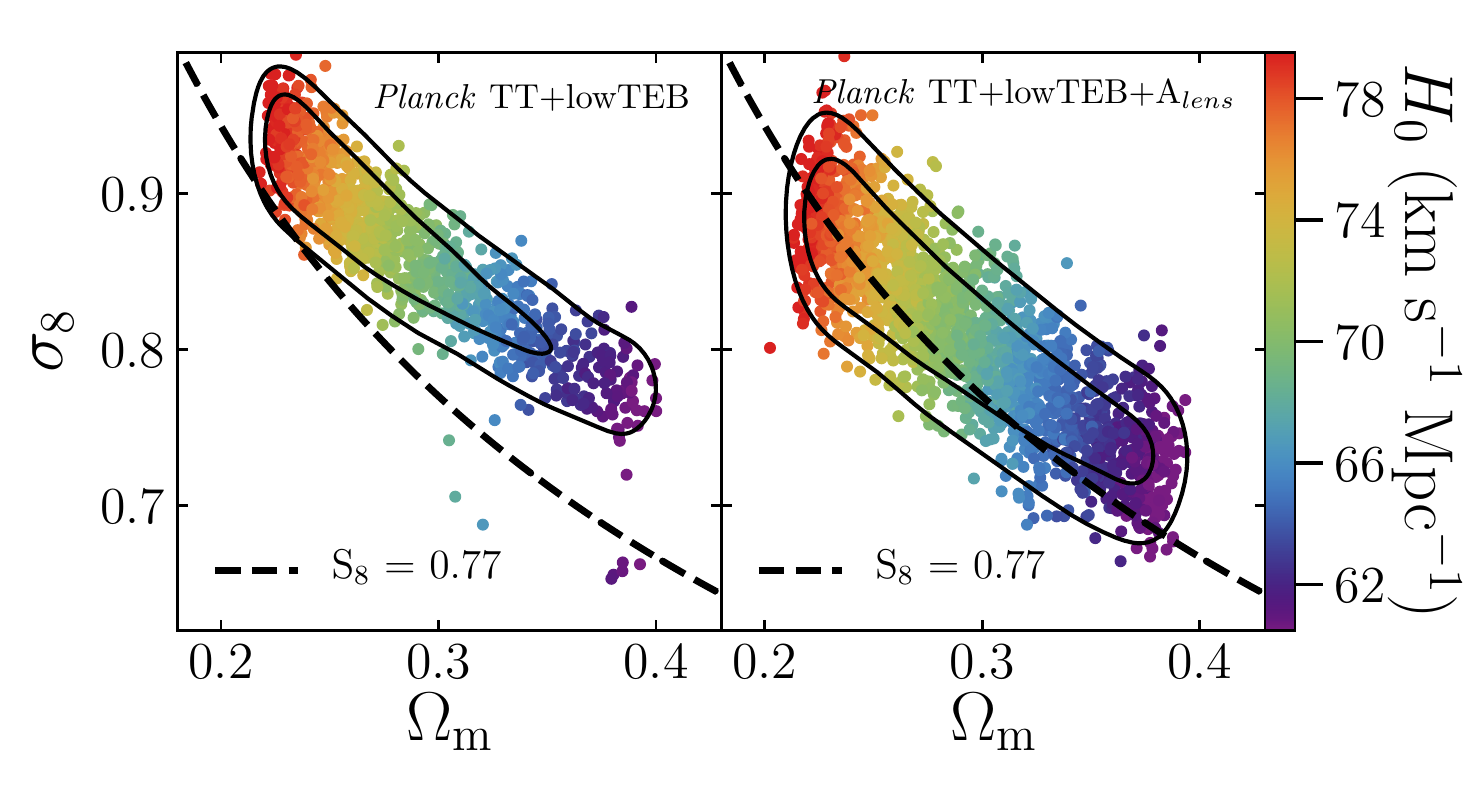}
    \caption{\textit{Top}: The 1$\sigma$ and 2$\sigma$ constraints in the $\textit{w}_0 - \textit{w}_{a}$ parameter space using Planck TT+lowTEB data, where A$_{\rm lens}$ has been fixed at unity (left column) or left to vary (right column). The black points show the locations of the simulated cosmologies and the error bars on the points show the size of the region used to generate the rest of the cosmological parameters. The dashed lines cross at the cosmological constant. \textit{Bottom}: The same as above except for $\Omega_{\rm m}-\sigma_8$, where the dashed line shows $S_8=0.77$.}
    \label{fig:alens}
\end{figure}

The \textit{Planck} collaboration has done extensive parameter estimations of $\Lambda$CDM and a variety of extensions, including DDE, with respect to the \textit{Planck} data and a combination of many other data sets \citep{planckparams2018}. This was done using \texttt{CosmoMC} \citep{antony2002} which is a Markov-Chain Monte-Carlo (MCMC) engine and a large quantity of the MCMC chains have been made public\footnote{The public chains are available from the  \href{www.wiki.cosmos.esa.int/planck-legacy-archive/index.php/Cosmological_Parameters}{Planck wiki}.}. However, the public library of MCMC chains are limited to only a few combinations of observational data for DDE cosmologies. Additionally, it is important to note the possibility of remaining systematics in the CMB data, one of which is the apparent enhanced smoothing of peaks and troughs in the temperature power spectrum. \citet{addison2016} have shown that this smoothing can be taken into account by letting the amplitude of the CMB lensing power spectrum, A$_{\rm lens}$, vary rather than setting it to unity (see also \citealt{calabrese2008,divalentino2016,renzi2018,mccarthy2018}). None of the publicly available chains for DDE include A$_{\rm lens}$ as a free parameter. Therefore we chose to use \texttt{CosmoMC} to generate our own chains as this gives us complete freedom over which parameters and observational data sets to include. Table~\ref{tab:priors} shows the parameters and their priors used with \texttt{CosmoMC}. All parameters with square brackets have uniform priors and single valued parameters were set to that constant. We used the data from the \textit{Planck} 2015 data release and the \texttt{GetDist} package \citep{getdist2019} to generate plots from the MCMC chains (see Fig~\ref{fig:simruns} and Fig~\ref{fig:alens}).

\begin{table}
	\centering
	\caption{The priors of the parameters used in the analysis with \texttt{CosmoMC}. Parameters with square brackets have uniform priors while single valued parameters are constants. From the top, the parameters are: baryon energy density, cold-dark-matter energy density, approximation to the observed angular size of the sound horizon at recombination, optical depth of reionisation, amplitude of scalar fluctuations, scalar spectral index, Hubble constant, two parameters defining the equation of state of dark energy (see Section~\ref{sec:dde}), sum of neutrino masses, effective number of relativistic degrees of freedom, and the amplitude of the CMB lensing power spectrum.}
	\begin{tabular}{lr}
		\hline
		Parameter & Prior\\
		\hline
		$\Omega_{\rm b}h^2$ & [0.005, 0.1]\\
		$\Omega_{\rm c}h^2$ & [0.001, 1.0]\\
		100$\theta_{\rm MC}$ & [0.5, 10.0]\\
		$\tau$ & [0.01, 0.8]\\
		ln($10^{10}A_{\rm s}$) & [2, 4]\\
		$n_{\rm s}$ & [0.8, 1.2]\\
		$H_0$ (km s$^{-1}$ Mpc$^{-1}$) & [60.0, 80.0]\\
		$\textit{w}_0$ & [-3.0, 1.0]\\
		$\textit{w}_{a}$ & [-3.0, 2.0]\\
		$\sum m_{\rm \nu}$ (eV) & 0.06\\
		$N_{\rm \nu}$ & 3.046\\
		$A_{\rm lens}$ & [0, 2]\\
		\hline
	\end{tabular}
    \label{tab:priors}
\end{table}

We first explore the $\textit{w}_0 - \textit{w}_{a}$ parameter space using a combination of the \textit{Planck} CMB temperature power spectrum (TT) and the polarisation power spectrum at low multipoles (lowTEB) \citep{plancklike}; a combination of BAO data from the SDSS Main Galaxy Sample \citep{ross2015}, the Baryon Oscillation Spectroscopic Survey (BOSS), BOSS CMASS and BOSS LOWZ \citep{anderson2014}, and the six-degree-Field Galaxy survey (6dFGS) \citep{beutler2011}; the supernova Ia constraints from the Joint Light-curve Analysis (JLA) data \citep{betoule2014}; and the constraints on $H_0$ from measurements of the local Universe \citep{riess2011}.

Fig.~\ref{fig:simruns} shows the 1$\sigma$ and 2$\sigma$ constraints in the $\textit{w}_0$-$\textit{w}_{a}$ parameter space for different combinations of data sets and for which A$_{\rm lens}=1$. The points are coloured by their $H_0$ value and the cosmological constant, $\textit{w}_0=-1$, $\textit{w}_{a}=0$, is indicated by the crossing of the dashed lines. The Planck TT+lowTEB data (top left) gives a broad contour with $H_0$ spanning a wide range of values that change in the direction perpendicular to the gradient of the contour. Adding BAO (top right) significantly reduces the allowed parameter space and limits the contour to lower values of $H_0$. Interestingly, neither of these contours are centered on the values of the cosmological constant, which sits at the boundary of the $1\sigma$ contour. The parameter space is further reduced along the degeneracy to a narrow region by adding JLA SNIa (bottom left). However, adding the local $H_0$ constraints instead of the JLA SNIa (bottom right) a much smaller effect on the allowed parameter space. These effects can be explained by the fact that the largest constraining power of the \textit{Planck} data on DDE comes from the distance to the surface of last scattering. Therefore, any expansion history is allowed as long as its integral returns the measured distance to the surface of last scattering. This geometric degeneracy within the $\textit{w}_0 - \textit{w}_{a}$ parameter space explains why the inclusion of BAO or type Ia supernovae significantly increases the constraining power on the $\textit{w}_0 - \textit{w}_{a}$ parameter space as they effectively probe the expansion history, $H(a)$.

Next, we explore the effect of A$_{\rm lens}$ on the allowed parameter space in Fig.~\ref{fig:alens}, which shows the $\textit{w}_0 - \textit{w}_{a}$ (top) and $\Omega_{\rm m}-\sigma_8$ (bottom) parameter spaces for the \textit{Planck} TT+lowTEB data with A$_{\rm lens}$ set to unity (left), as done in the \textit{Planck} analysis, and as a free parameter (right). For reference, the LSS joint constraint S$_8=\sigma_8\sqrt{\Omega_{\rm m}/0.3}=0.77$ is shown on the $\Omega_{\rm m}-\sigma_8$ plot (dashed line).

Including $A_{\rm lens}$ as a free parameter stretches the contour of the $\textit{w}_0 - \textit{w}_{a}$ parameter space towards lower (higher) values of $\textit{w}_0$ ($\textit{w}_{a}$). It is interesting to note that the cosmological constant, $\textit{w}_0=-1$ and $\textit{w}_{a}=0$, is in mild tension with \textit{Planck} if A$_{\rm lens}$ is fixed at unity, the default value adopted by \textit{Planck}, but reconciled if it is allowed to vary. For the bottom of Fig.~\ref{fig:alens}, leaving $A_{\rm lens}$ as a free parameter systematically shifts the contour to lower values of $\sigma_8$, resulting in a much better agreement with the LSS joint constraint. 

\begin{table*}
    \centering
    \caption{The cosmological parameters of our 6 chosen cosmologies derived from the $Planck$ CMB data (TT+lowTEB) with marginalisation over the lensing amplitude, $A_{\rm lens}$. From left to right, the parameters are: (1) and (2) the 2 free parameters describing DDE (see Equation~\ref{equ:deeos}), (3) the total matter density at present-day, (4) the baryon density at present-day, (5) Hubble's constant, (6) the spectral index of the initial power spectrum, (7) the amplitude of the power spectrum at recombination at a pivot scale of 0.05 Mpc$^{-1}$, (8) the optical depth to reionization, (9) the amplitude of the linear matter power spectrum on 8 Mpc/h scales at present-day, (10) S$_8=\sigma_8\sqrt{\Omega_{\rm m}/0.3}$, (11) the amplitude of the CMB lensing power spectrum.}
	\begin{tabular}{ccccccccccc}
	    \hline
	    (1) & (2) & (3) & (4) & (5) & (6) & (7) & (8) & (9) & (10) & (11)\\
		\hline
		$\textit{w}_0$ & $\textit{w}_{a}$ & $\Omega_{\rm m}$ & $\Omega_{\rm b}$ & $H_0$ & $n_{\rm s}$ & $A_{\rm s}$ & $\tau$ & $\sigma_8$ & $S_8$ & $A_{\rm lens}$\\
		 & & & & (km/s/Mpc) & & (10$^{-9}$) & & & & \\
		\hline
        -1.16 & 0.73 & 0.309 & 0.0501 & 67.25 & 0.975 & 2.10 & 0.058 & 0.773 & 0.783 & 1.298\\
        -1.00 & 0.00 & 0.294 & 0.0476 & 68.98 & 0.974 & 2.11 & 0.061 & 0.802 & 0.795 & 1.233\\
        -0.84 & -0.73 & 0.288 & 0.0465 & 69.73 & 0.974 & 2.11 & 0.060 & 0.815 & 0.798 & 1.205\\
        -0.67 & -1.45 & 0.286 & 0.0462 & 69.97 & 0.973 & 2.10 & 0.059 & 0.819 & 0.801 & 1.195\\
        -0.51 & -2.18 & 0.284 & 0.0459 & 70.20 & 0.974 & 2.10 & 0.060 & 0.822 & 0.800 & 1.194\\
        -0.35 & -2.89 & 0.289 & 0.0465 & 69.71 & 0.973 & 2.10 & 0.059 & 0.824 & 0.806 & 1.174\\
		\hline
	\end{tabular}
	\label{tab:cosmoparams}
\end{table*}

In order to generate our cosmologies for the simulations, we sampled the geometric degeneracy in the $\textit{w}_0 - \textit{w}_{a}$ parameter space shown in Fig.~\ref{fig:alens} that includes A$_{\rm lens}$ as a free parameter. We opted not to use data sets other than the CMB to further constrain this parameter space, for three reasons: i) as discussed in the introduction, there are known tensions between `early' (CMB+BAO) and `late' ($H_0$) Universe measures\footnote{Type Ia supernovae constraints can agree with either, depending on how the distance scale to supernovae is established (i.e., via Cepheids or BAO with a CMB-based estimate of the physical sound horizon) \citep{macaulay2019}.} of the expansion history, making the combination of these constraints questionable; ii) the CMB-only (without BAO) constraints are fully compatible with any of the possible data set combinations; and iii) the CMB-only constraints allow for the largest variation in DDE models, resulting in a wider range of behaviours to study from a theoretical perspective.

We choose 6 equally spaced points along the degeneracy to get 6 values of $\textit{w}_0$ and $\textit{w}_{a}$, one of which is the cosmological constant and is referred to as the reference $\Lambda$CDM cosmology throughout. To specify the other cosmological parameters for each choice of $\textit{w}_0$ and $\textit{w}_{a}$, we calculate the weighted average of each parameter from every sample of the MCMC chain that contain the values of $\textit{w}_0\pm0.05$ and $\textit{w}_{a}\pm0.05$. In this way, all of the simulations are guaranteed to be compatible with the primary CMB angular power spectrum.  The resulting 6 cosmologies are listed in Table~\ref{tab:cosmoparams}. All of our cosmologies are spatially flat, i.e. $\Omega_{\rm k}=0$.

\begin{figure}
	\includegraphics[width=\columnwidth]{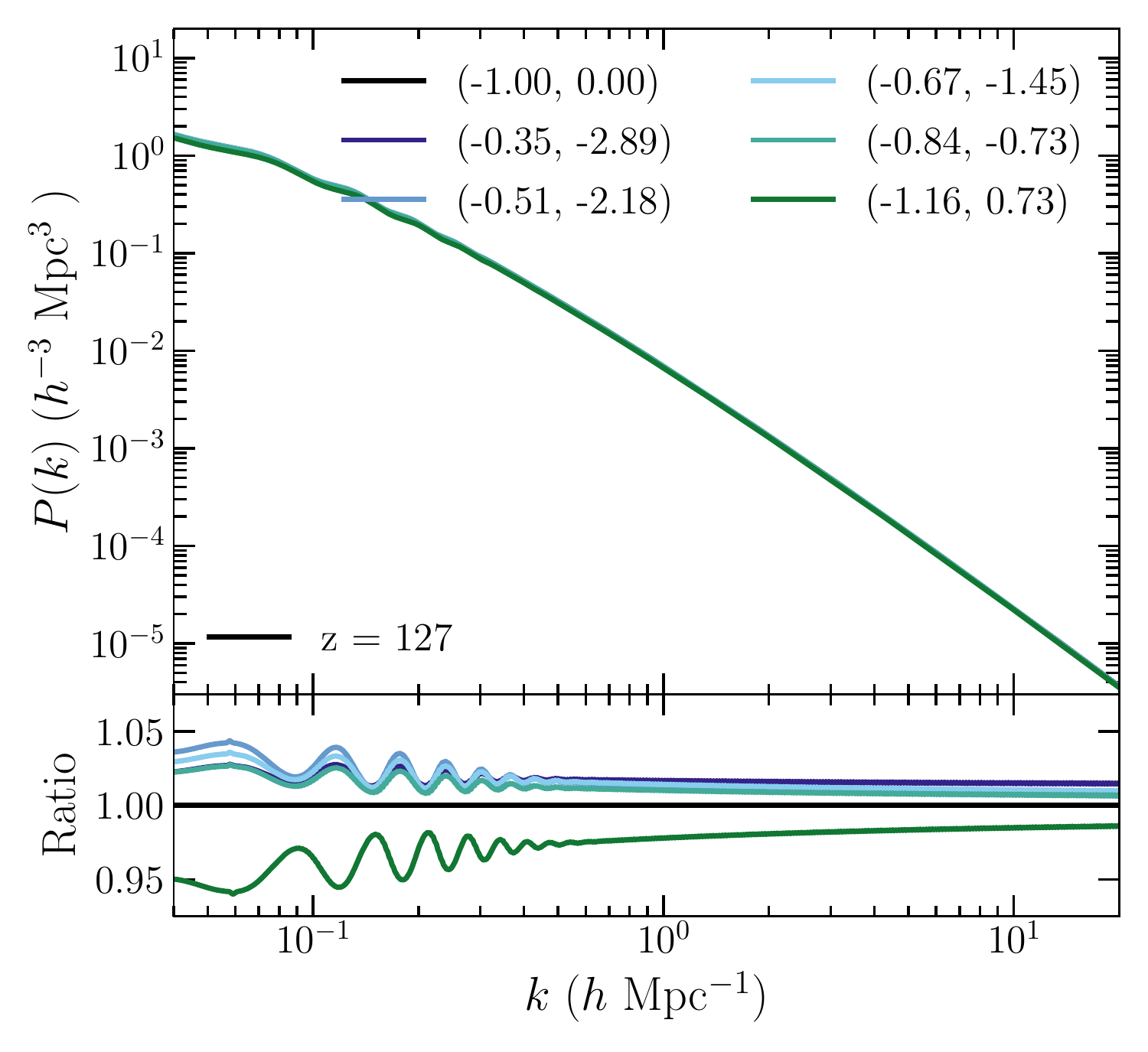}
    \caption{\textit{Top}: The matter power spectrum of the ICs for each cosmology at $z=127$ computed with \texttt{CAMB}. \textit{Bottom}: The ratios of matter power spectra relative to $\Lambda$CDM. Colours indicate different cosmologies where bracketed values refer to the values of ($\textit{w}_0, \textit{w}_{a}$).}
    \label{fig:matterpowerspec127}
\end{figure}
We plot the matter power spectra of the ICs for each cosmology in Fig.~\ref{fig:matterpowerspec127} to show that these cosmologies already have different matter distributions at high redshift. The power spectra were generated using \texttt{CAMB} at the simulation starting redshift of $z=127$. The cosmologies already have a difference of $\approx5\%$ in $P(k)$ at large scales (small $k$) and $\approx1\%$ at small scales (large $k$) before starting the simulations. Due to slight offsets in the power spectra, the BAO signal at $k\sim0.1$ becomes apparent in the ratios.

\begin{figure}
	\includegraphics[width=\columnwidth]{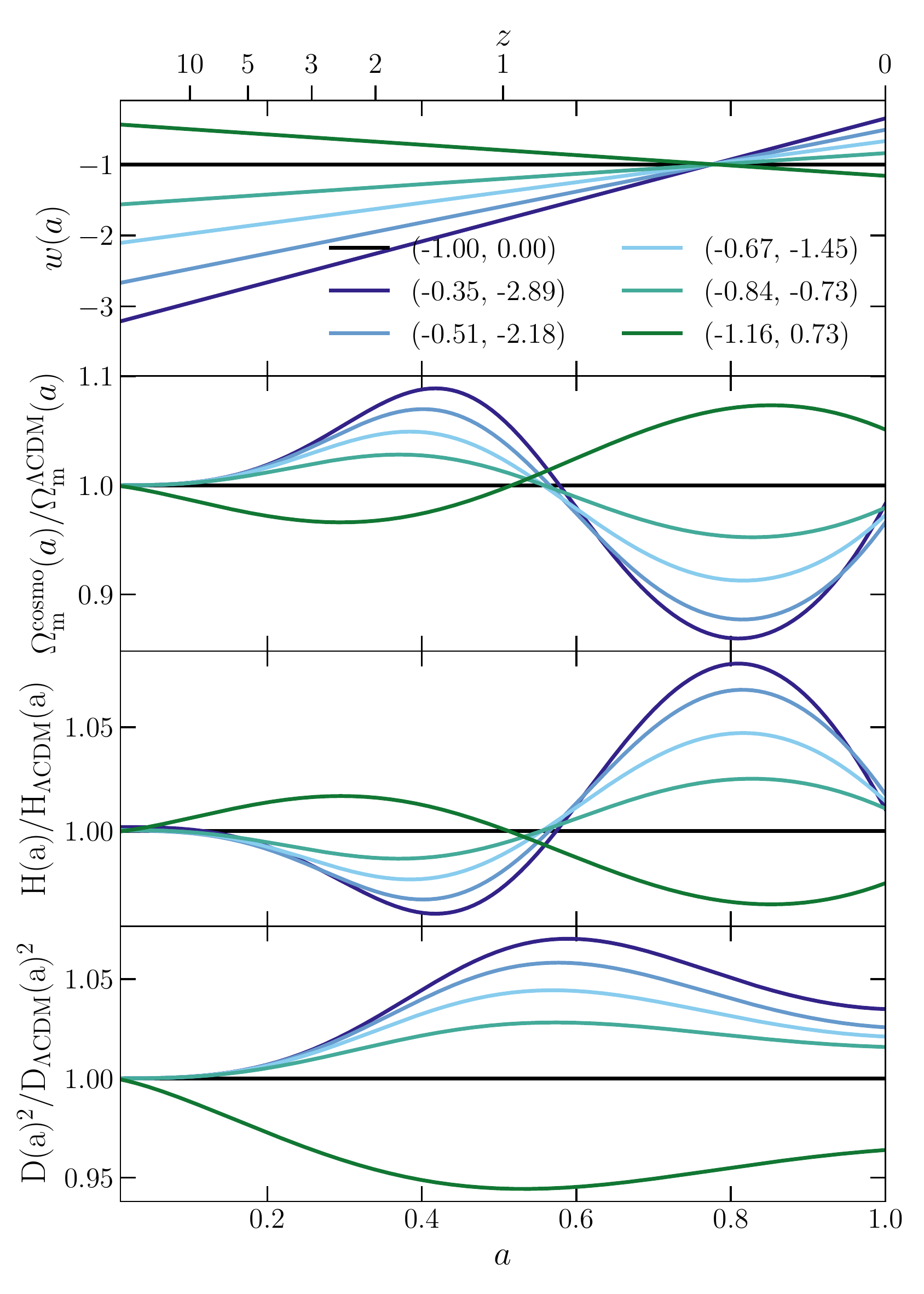}
    \caption{The evolution of $\textit{w}(a)$ given by Equation~\ref{equ:deeos} (top), $\Omega_{\rm m}$ (middle-top), expansion history (middle-bottom) and linear growth factor (bottom) as a function of expansion factor and redshift for the cosmologies shown in Table \ref{tab:cosmoparams}. Each statistic, apart from $\textit{w}(a)$, has been normalised by the $\Lambda$CDM cosmology. Colours indicate different cosmologies where bracketed values refer to the values of ($\textit{w}_0, \textit{w}_{a}$).}
    \label{fig:expansionhistory}
\end{figure}

Our DDE terminology is based on quintessence models which can be classified into two categories: `thawing' models start at $\textit{w}\approx-1$ and have $\textit{w}(a)$ increase with $a$ \citep{caldwell2005,scherrer2008,chiba2009,gupta2015}, whereas `freezing' models have $\textit{w}(a)$ decrease with $a$ and approach $\textit{w}\approx-1$ at late times \citep{caldwell2005,scherrer2006,chiba2006,sahlen2007}. We will adopt this terminology throughout, calling models with $\textit{w}_{a}<0$ thawing and $\textit{w}_{a}>0$ freezing, although we note that our models can cross the $\textit{w}=-1$ threshold, which is not the case for quintessence models. The evolution of $\textit{w}(a)$ is shown in the top panel of Fig.~\ref{fig:expansionhistory}, where the line above $\textit{w}=-1$ is our freezing cosmology and the lines below are our 4 thawing cosmologies.

Now that we have selected the cosmologies, it is possible to examine some useful physical quantities before running any simulations (these will be useful for interpreting the simulation-based results later).
Fig.~\ref{fig:expansionhistory} also shows the evolution of $\Omega_{\rm m}(a)$ (middle top) and $H(a)$ (middle bottom) for the different cosmologies, normalised by the $\Lambda$CDM cosmology. These have been calculated using Equation~\ref{expansionhistory}. We also show the linear growth factor, $D(a)$, for each cosmology normalised by the $\Lambda$CDM cosmology (bottom). The linear growth factor is defined as the ratio of matter overdensities at a given scale factor, $\delta(a)$, relative to some initial overdensity, $D(a)=\delta(a)/\delta(a_i)$. The closed form approximation \citep{peebles1980,eisenstein1997} typically used to calculate $D(a)$ is valid for $\Lambda$CDM but does not return the correct results for DDE cosmologies. Instead, equations such as those presented in \citet{linder2003growth} should be solved.

It is clear from Fig.~\ref{fig:expansionhistory} that the thawing dark energy models behave systematically different to the freezing model. Any general trend in the former is the inverse in the latter. The largest differences appear at $z<1$, as one might expect since dark energy dominates the energy density of the Universe at late times.  All of our models cross at the same $\textit{w}(a)$ and $a$ (top of Fig.~\ref{fig:expansionhistory}) because of the way we choose our cosmological models. To show why this is, one can equate Equation~\ref{equ:deeos} for two different models [e.g. ($\textit{w}_{0,1}$, $\textit{w}_{a,1}$) and ($\textit{w}_{0,2}$, $\textit{w}_{a,2}$)] and solve for expansion factor, $a$, at which $w(a)_1 = w(a)_2$:
\begin{equation}
    a=1+\frac{\textit{w}_{0,2} - \textit{w}_{0,1}}{\textit{w}_{a,2}-\textit{w}_{a,1}}=1+\frac{\rm d\textit{w}_{0}}{\rm d\textit{w}_{a}}.
    \label{equ:waw0grad}
\end{equation}
\noindent Equation~\ref{equ:waw0grad} shows that any DDE models that lie on the same line in the $\textit{w}_0 - \textit{w}_{a}$ parameter space (which is the case here, as we select values along the CMB geometric degeneracy) will all cross at the same value of $a$, with that value depending only on the slope of the line.  This feature, along with the fact that the line corresponds to a geometric degeneracy (i.e., the models are all constrained to yield the same distance to the last-scattering surface), is also likely responsible for the similar scale factors at which $\Omega_{\rm m}(a)$ and $H(a)$ cross. 

\section{Large-scale structure}
\label{sec:lss}

In this section we explore the impact of our DDE cosmologies on a number of common measures of LSS, including the matter power spectrum ($P(k)$), the halo 2-point auto-correlation function, the halo mass function and halo number counts. We use the collisionless (dark matter-only) versions of the simulations (the impact of baryons is discussed in Section \ref{sec:separability}).  We discuss how the DDE cosmologies affect these LSS statistics and draw comparisons with other cosmologies constrained by the CMB which we explored in previous BAHAMAS papers; the effects of massive neutrinos \citep{mummery2017} and running of the spectral index \citep{stafford2020}.

\subsection{Matter power spectrum}
\label{sec:matterclustering}

\begin{figure}
	\includegraphics[width=\columnwidth]{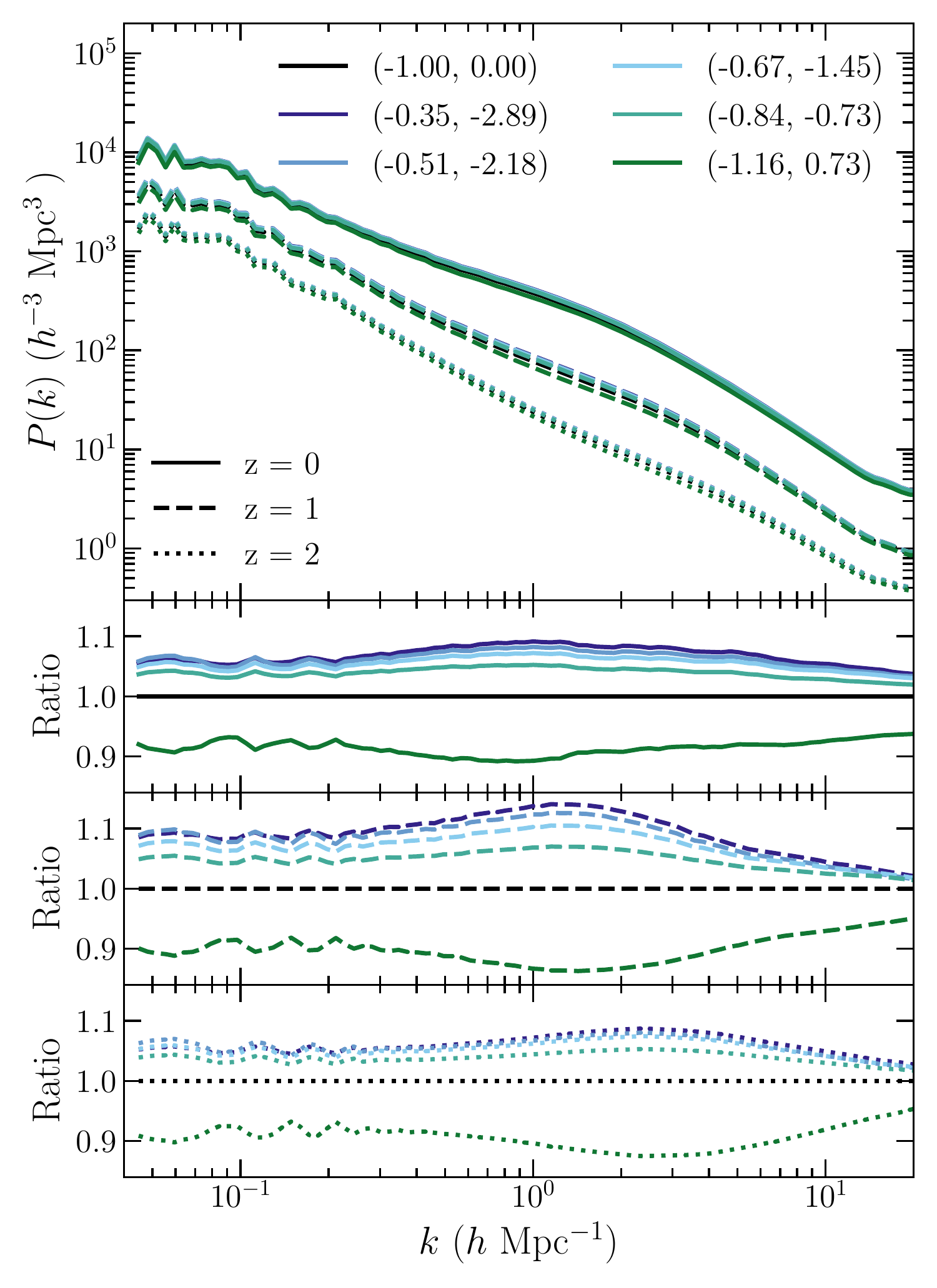}
    \caption{\textit{Top}: The total matter power spectrum of the collissionless simulations for the different cosmologies and redshifts. \textit{Bottom}: The ratios of matter power spectra relative to $\Lambda$CDM at each redshift. Colours indicate different cosmologies where bracketed values refer to the values of ($\textit{w}_0, \textit{w}_{a}$) while line styles show redshift.}
    \label{fig:matterpowerspec}
\end{figure}

We first investigate the effect of our DDE cosmologies on the matter clustering via the non-linear matter power spectrum of the total matter in our collisionless simulations. The power spectra are computed using the \texttt{GenPK}\footnote{https://github.com/sbird/GenPK} code \citep{bird2017}.

Fig.~\ref{fig:matterpowerspec} shows the total matter power spectrum of the collisionless simulations for the different cosmologies at $z=0,1,2$, where ratios have been taken with respect to the $\Lambda$CDM cosmology. Since we used the same phases to generate the ICs for each cosmology, we do not need to worry about cosmic variance issues and the ratio of $P(k)$ between two different simulations should be an accurate and robust prediction.

The freezing dark energy model shows a suppression in power of $\approx$10\%, whereas the thawing dark energy models show an increase in power of $\approx$5-10\%. This effect is slightly scale dependent with maximum impact at $k\approx1$ $h$ Mpc$^{-1}$ and the largest change in $P(k)$ is seen at $z=1$.  The change in amplitude and the redshift evolution of $P(k)$ on linear scales (i.e., low $k$ values) agrees with naive expectations based on the behaviour of  $D(a)$ in Fig.~\ref{fig:expansionhistory}. Note that the amplitude of $P(k)\propto D^2(a)$ in the linear regime. While the use of $D(a)$ is only strictly valid on linear scales, it is interesting to note that the change to $P(k)$ from DDE propagates through to non-linear scales. This can be explained through `mode mixing', where $k$-modes no longer evolve independently from each other, but transfer power from large to small scales.

One can compare these effects to alternative extensions to the $\Lambda$CDM cosmology. \citet{mummery2017} (hereafter M17) examined massive neutrino extensions and found that neutrinos suppress the matter power spectrum between $\approx$5\% and $\approx$30\% for the lowest, $\Sigma M_{\nu}=0.06$ eV, and largest sum of neutrino masses, $\Sigma M_{\nu}=0.48$ eV, respectively.
Interestingly, the suppression in $P(k)$ from massive neutrinos has a similar shape to the DDE cosmologies in Fig.~\ref{fig:matterpowerspec}, which could act to mask a combination of massive neutrinos and DDE.  Another possible extension to $\Lambda$CDM is the inclusion of a running of the scalar spectral index, $n_{\rm s}$, which was investigated recently by \citet{stafford2020} (hereafter S20).  They found that negative (positive) running results in an amplification (suppression) of the matter power spectrum of $\approx$5-10\%. These effects had a scale dependence that caused a decrease in their magnitude towards smaller scales, especially at higher redshifts. 

In addition, it is well known that baryonic effects on the matter power spectrum are of the order of $\sim$10-20\% and cause a suppression in the power spectrum at k$\ga$0.1 Mpc$^{-1} h$ \citep{vandaalen2011,mummery2017,schneider2019,vandaalen2020, debackere2020}. The DDE cosmologies considered here produce effects of similar magnitude, although they extend throughout the linear and non-linear regime and should therefore be distinguishable from baryonic effects given a wide enough range of well-sampled $k$ values. We explore this in Section \ref{sec:separability}.

\subsection{Halo clustering}
\label{sec:haloclustering}

\begin{figure}
	\includegraphics[width=\columnwidth]{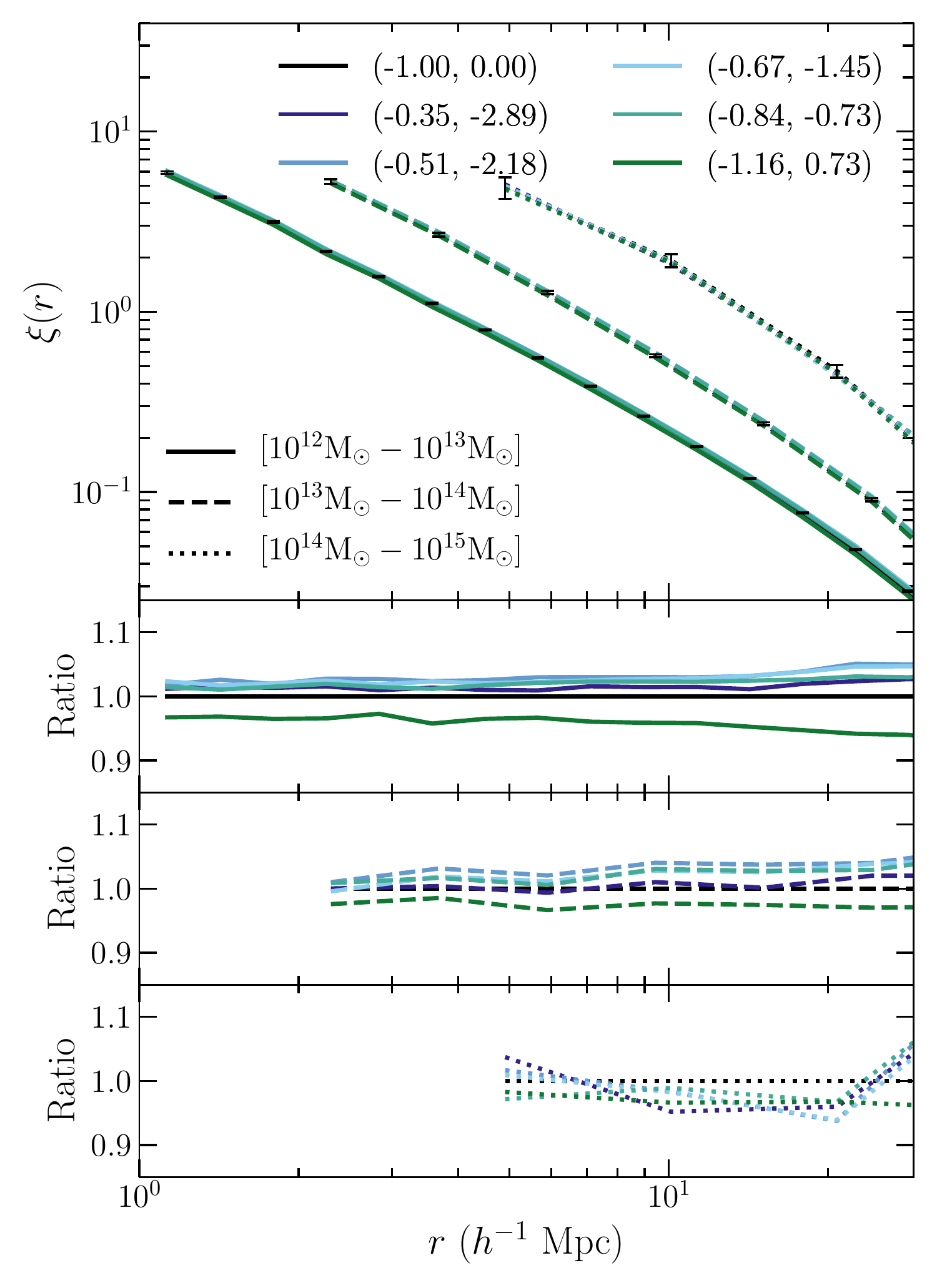}
    \caption{\textit{Top}: The 2-point auto-correlation function of dark matter haloes for the different cosmologies and mass bins at $z=0$. \textit{Bottom}: The ratios of the 2-point correlation functions relative to the $\Lambda$CDM cosmology at different redshifts Colours indicate different cosmologies where bracketed values refer to the values of ($\textit{w}_0, \textit{w}_{a}$) and line styles show separate mass bins given in M$_{200,\rm crit}$. The cut-off at small radii is due to the overlapping of haloes which forces $\xi$ to turn over. Error bars represent the Poisson uncertainties determined from the number of haloes in each radial bin for the $\Lambda$CDM cosmology.}
    \label{fig:2pcf}
\end{figure}

The clustering of dark matter haloes can be described by the 2-point auto-correlation function, $\xi(r)$, which is the excess probability of finding two haloes with a given separation, $r$, relative to a random distribution of haloes \citep{davis1983}. To compute this, one calculates the separation, $r$, between each halo and every other halo in the sample. The distribution of halo separations in bins of $r$ can then be defined as $DD(r)$. The separation pair count of a random distribution, $RR(r)$, can be calculated analytically assuming the halos are distributed homogeneously with a density equal to the total number of haloes in the sample divided by the volume of the simulation. The 2-point auto-correlation function is then
\begin{equation}\label{2pcf}\centering
    \xi(r) = \frac{DD(r)}{RR(r)} - 1.
\end{equation}
Fig.~\ref{fig:2pcf} shows the 2-point auto-correlation function for dark matter haloes in three mass bins of M$_{200, \rm crit}$. The ratios are shown relative to the $\Lambda$CDM cosmology. In general, the freezing (thawing) dark energy cosmology produces haloes with decreased (increased) clustering relative to $\Lambda$CDM, generally mimicking the behaviour in $P(k)$. The lowest mass bin shows a $\approx$10\% effect which decreases towards higher masses. Haloes start to overlap on small scales causing the 2-point auto-correlation function turn over and decrease which is where we introduce a cut-off. As the size of haloes increases with increasing mass, this cut-off shifts to larger radii. We show the statistical errors on the 2-point auto-correlation function for the $\Lambda$CDM cosmology which were taken to be the Poisson uncertainties on the number of haloes in each radial bin. The errors for the other cosmologies are approximately equal to those of the $\Lambda$CDM cosmology. The uncertainties are slightly larger in bins at lower radii (as they sample smaller volumes) and for higher masses due to their lower abundance. Since we use the same phases to generate the ICs, we can compare the ratios between the different cosmologies without the complication of cosmic variance. That also means that measurements between simulations are strongly correlated. Therefore we only show the Poisson error on the absolute value and not in the lower ratio panels. 

This change in the clustering of haloes is analogous to the change in the matter power spectrum, $P(k)$ seen in Fig.~\ref{fig:matterpowerspec}, which is unsurprising since the 2-point auto-correlation function is the Fourier transfer of $P(k)$ multiplied by the linear halo bias, $b^2$.

The 2-point auto-correlation was also calculated for matched haloes. Matching haloes is done by identifying the 50 most bound dark matter particles comprising a halo in the $\Lambda$CDM simulation using their unique particle IDs and finding the halo in another simulation that contains the majority of dark matter particles with the same IDs. By inspecting a set of matched haloes we remove any additional effect due to the change in halo mass for different cosmologies, as seen in Section~\ref{sec:hmf} below. The general trends of the 2-point auto-correlation function for matched haloes is the same as for unmatched haloes, although with increased effect due to the change in halo mass between different cosmologies. This is due to the fact that more massive haloes are more biased tracers of the underlying matter clustering and therefore show a higher clustering signal in the 2-point auto-correlation function.

M17 finds that massive neutrinos suppress the 2-point auto-correlation function of haloes with M$_{200, \rm crit}$=$10^{12}$M$_{\odot}$-$10^{13}$M$_{\odot}$ by $\approx$5\% and $\approx$20\% for the lowest and largest sum of neutrino masses, respectively. S20 shows that their cosmologies with running of the spectral index enhances the clustering signal by $\approx$5\% for negative running and vice versa for positive running for haloes within the same mass range. This is very similar to the effects of DDE which, unlike massive neutrinos, cannot only suppress but also enhance the clustering signal relative to the $\Lambda$CDM cosmology.

\subsection{Halo mass function}
\label{sec:hmf}
The first statistic of halo abundance we examine is the halo mass function (HMF), $\Phi$, defined as the number of haloes per comoving volume per logarithmic unit of mass M$_{200,\rm crit}$,

\begin{equation}
    \Phi \equiv \frac{dn}{d\log_{10}(\rm M_{200,\rm crit})}.
\end{equation}{}

\begin{figure}
	\includegraphics[width=\columnwidth]{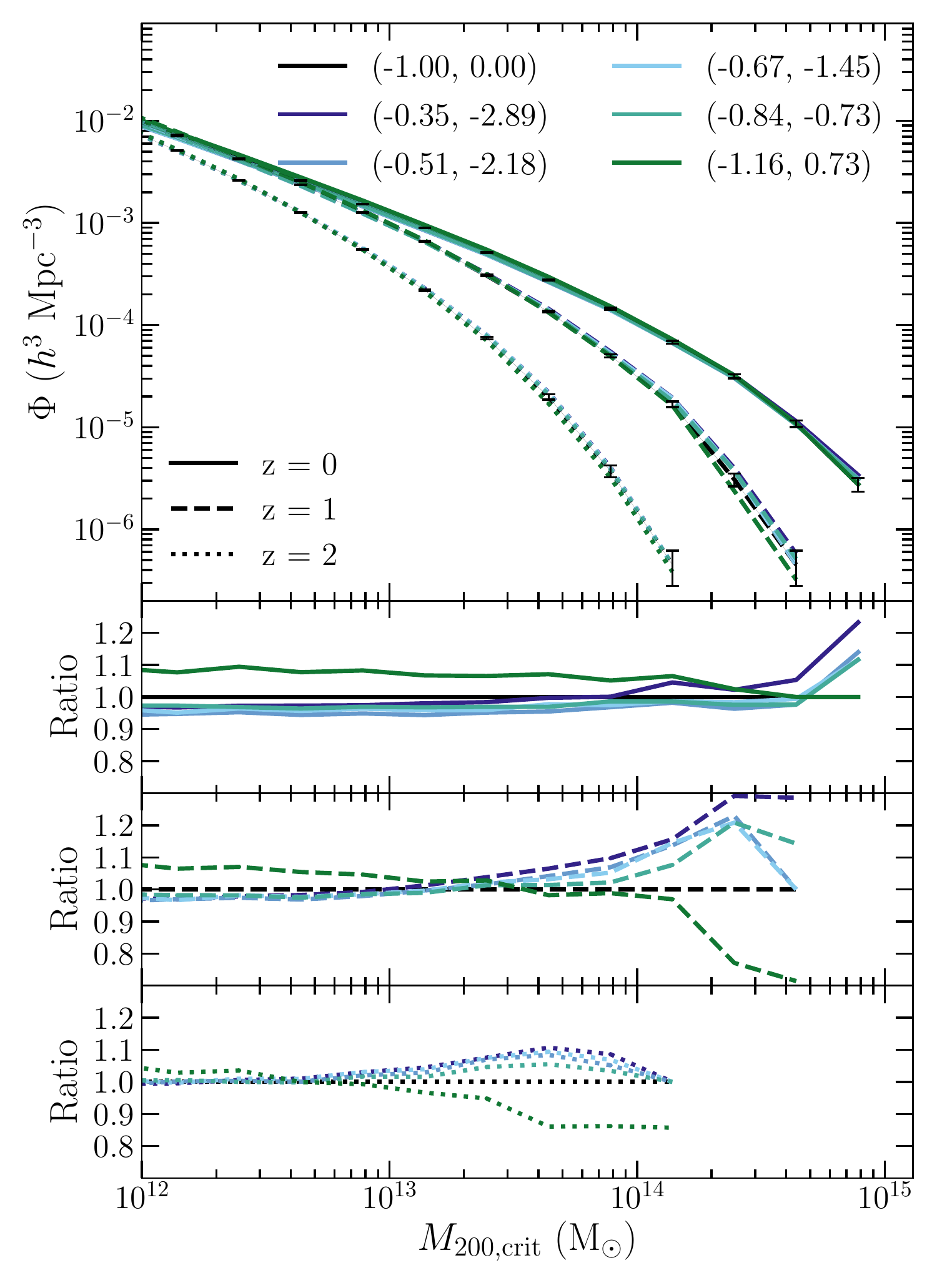}
    \caption{\textit{Top}: The HMF of the collisionless (dark matter-only) simulations for the different cosmologies and redshifts. \textit{Bottom}: The ratios of the HMFs with respect to the $\Lambda$CDM cosmology for each redshift. Colours indicate different cosmologies where bracketed values refer to the values of ($\textit{w}_0, \textit{w}_{a}$) and line styles indicate different redshifts. Error bars represent the Poisson uncertainties from the number of haloes in each mass bin for the $\Lambda$CDM cosmology.}
    \label{fig:hmf}
\end{figure}

In Fig.~\ref{fig:hmf} we show the HMF for the collisionless simulations of the different cosmologies at different redshifts, where the ratios are with respect to the $\Lambda$CDM cosmology. At $z=0$, the freezing dark energy model has a higher (lower) number density of low-mass (high-mass) haloes, while for the thawing models this trend is reversed. These effects are most apparent at $z=1$ where a change in the abundance of high-mass haloes of $\sim$20\% is seen and a crossover appears in the ratios at M$_{200,\rm crit}\sim$10$^{13}$M$_\odot$. The behaviour of the HMF is very different to that of $P(k)$, which shows no crossover and the opposite behaviour to the effect seen on low masses for the HMF.  We show the statistical errors on the HMF for the $\Lambda$CDM cosmology which were taken as the Poisson uncertainties from the number of haloes in each mass bin. The errors for the other cosmologies are approximately equal to those of the $\Lambda$CDM cosmology. The uncertainties are significant at the highest masses due to the rarity of such haloes in our simulations.

\begin{figure}
	\includegraphics[width=\columnwidth]{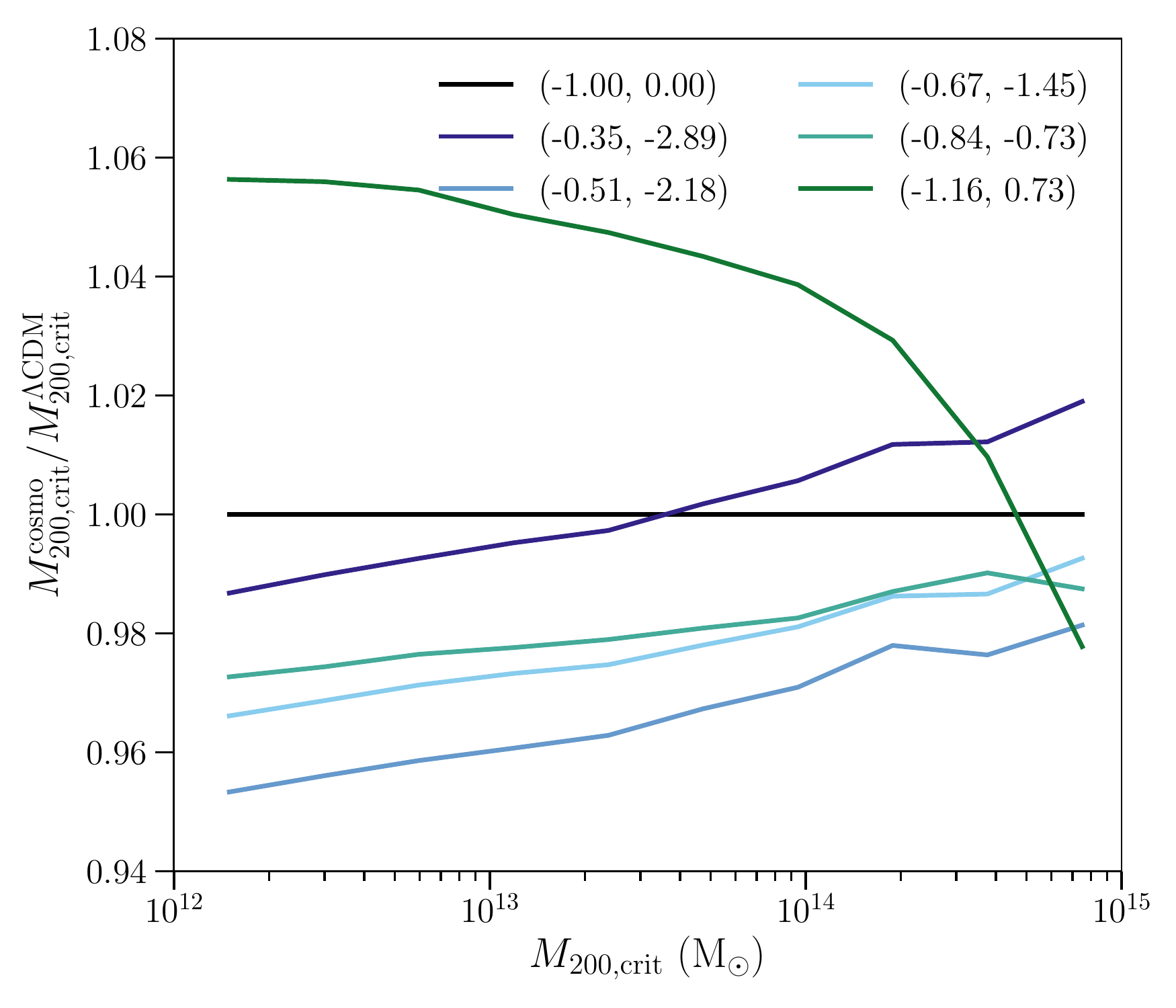}
    \caption{The median fractional change in halo mass relative to matched haloes from the $\Lambda$CDM cosmology at $z=0$. All haloes have been matched to the $\Lambda$CDM cosmology. Colours indicate different cosmologies where bracketed values refer to the values of ($\textit{w}_0, \textit{w}_{a}$).}
    \label{fig:massfraction}
\end{figure}

Another way of looking at this effect is to plot the change in halo mass between matched haloes from different cosmologies rather than halo abundance. Fig.~\ref{fig:massfraction} shows the fractional change in halo mass relative to matched haloes from the $\Lambda$CDM cosmology at $z=0$. This is plotted against the halo mass of the matched halo from the $\Lambda$CDM cosmology. Here we look at the change in halo mass at fixed abundance rather than changes in abundance at fixed halo mass. In this format, a vertical change in the fractional halo mass is comparable to a horizontal shift in the HMF. The trends in the HMF are also seen in the fractional change in halo mass with similar amplitude and mass scale. The freezing DDE cosmology forms more massive low-mass haloes but the growth of structure is suppressed and so the most massive haloes are not as massive as their $\Lambda$CDM equivalent. This trend is reversed for the thawing DDE cosmologies.

We can decompose the difference in the HMF between the different cosmologies into two effects. Firstly, the almost constant offset in the ratios of the HMF at the low-mass end (most apparent at $z=0$) can be explained by the difference in $\Omega_{\rm m}$ for the different cosmologies because dark matter haloes grow more massive in a cosmology with a higher $\Omega_{\rm m}$. Secondly, the crossover in the ratios at the high-mass end is due to the change in the growth of structure that is also seen in $P(k)$ in Fig.~\ref{fig:matterpowerspec}. The freezing cosmology shows a suppression in the growth of structure through the suppression in $P(k)$, meaning that high-mass haloes, which are still collapsing at that time, are less abundant with respect to the $\Lambda$CDM cosmology. This concept is explored further using the HMF fitting function of \citet{tinker2008} in Appendix~\ref{sec:appendixa}.

M17 showed that massive neutrinos suppress the HMF with the largest effect at the high-mass end. Halo masses are suppressed by $\approx$10\% and $\approx$50\% for the lowest and largest sum of neutrino masses, respectively. Interestingly, S20 found that cosmologies which include running of the spectral index can impact the HMF in a very similar way to the DDE cosmologies, suppressing/amplifying the HMF at low/high masses for negative running cosmologies and vice versa for positive running cosmologies.

The effects of DDE on the HMF are very different to the effects of baryons on the HMF over the masses sampled here.  M17 showed that baryonic feedback tends to suppress the HMF more strongly towards the high-mass end. However, at very high masses the gravitational potential is strong enough to counteract the feedback, thus reducing its effect on the HMF.  As well as this mass dependence, the amplitude of the baryonic impact is much stronger than that of the DDE cosmologies when they are constrained to reproduce the primary CMB (particularly the angular scale of the acoustic peaks).

\begin{figure}
	\includegraphics[width=\columnwidth]{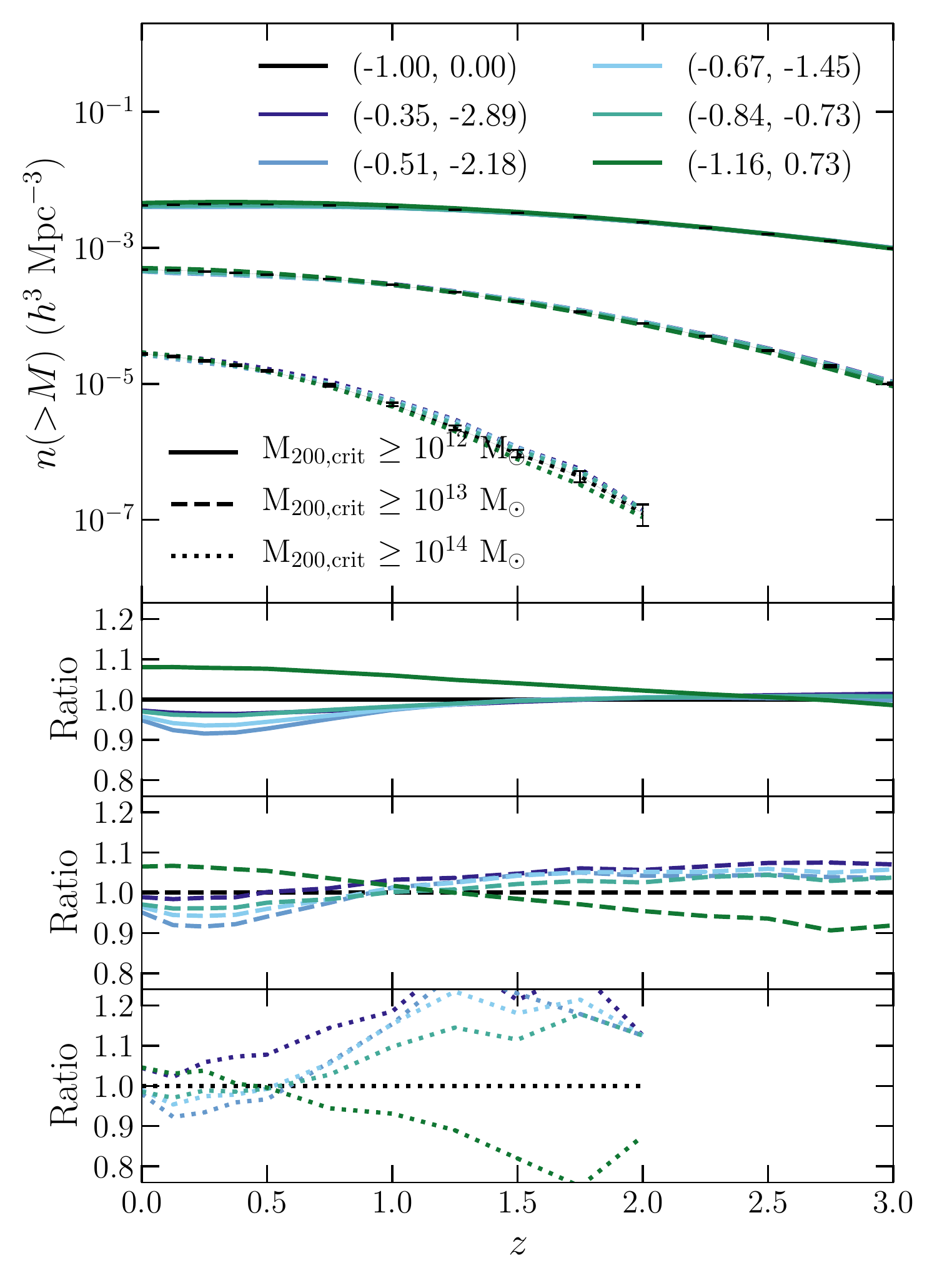}
    \caption{\textit{Top}: The number density of dark matter haloes for different cosmologies and mass cuts. \textit{Bottom}: The ratios of number density relative to the $\Lambda$CDM cosmology. Colours indicate different cosmologies where bracketed values refer to the values of ($\textit{w}_0, \textit{w}_{a}$) and the line styles show different lower mass limits of 10$^{12}$M$_\odot$, 10$^{13}$M$_\odot$ and 10$^{14}$M$_\odot$. The error bars represent the Poisson uncertainties derived from the number of haloes in each redshift bin for the $\Lambda$CDM cosmology.}
    \label{fig:clustercounts}
\end{figure}

\subsection{Halo number counts}
\label{sec:cc}
Next we examine the halo space density at a given redshift computed by integrating the HMF above a given mass. The halo space density simply represents the number density of haloes above a given mass. This is similar to what is more typically measured observationally since many surveys have too small of a volume to robustly measure the HMF, especially at high masses. 

\begin{figure*}
	\includegraphics[width=\textwidth]{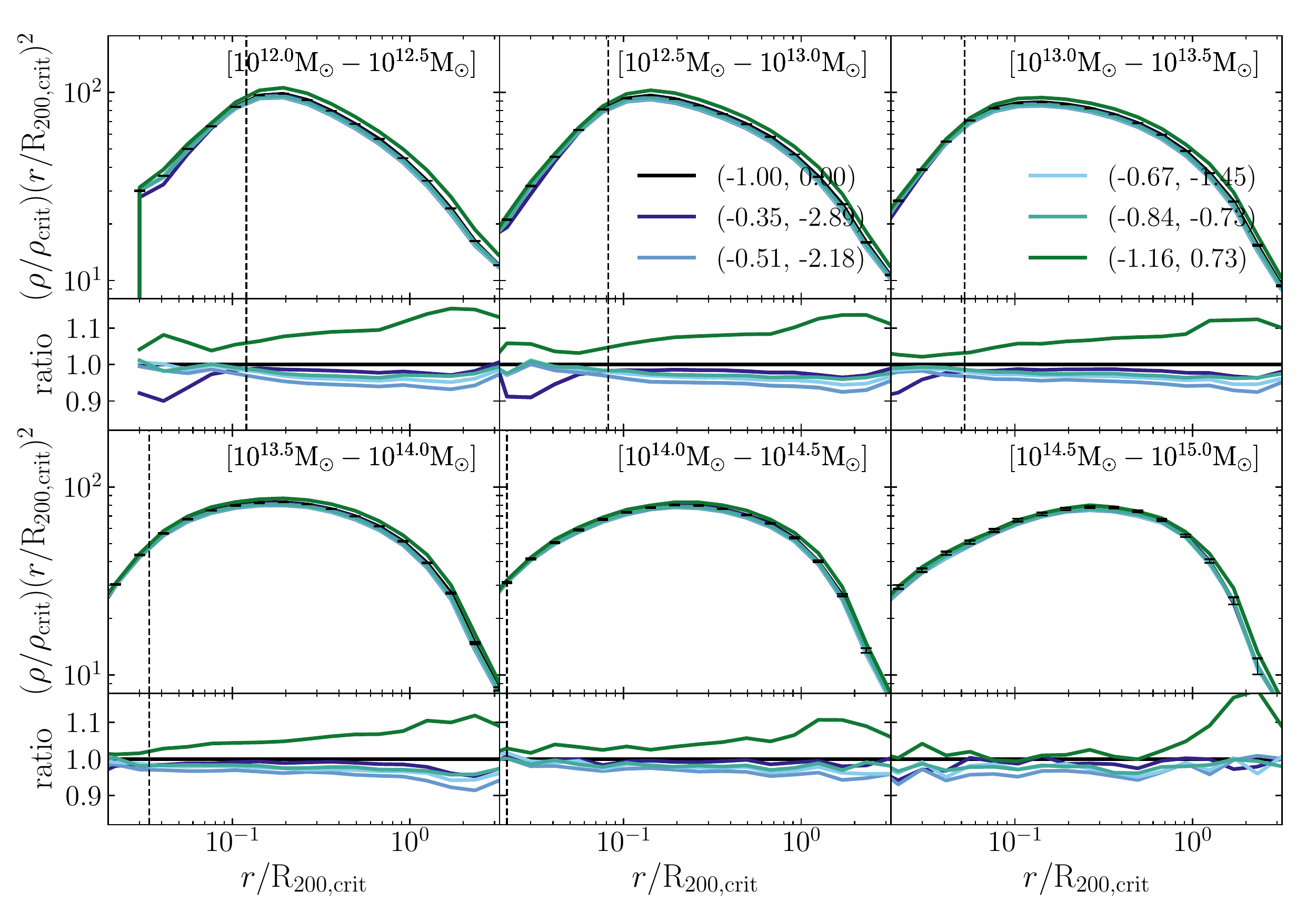}
    \caption{Median radial total mass density profiles of matched haloes for the different DDE cosmologies for collisionless simulations. Haloes have been matched to the $\Lambda$CDM cosmology. The panels show different mass bins with width 0.5 dex in M$_{200,{\rm crit}}$ for the $\Lambda$CDM cosmology. Colours indicate different cosmologies where bracketed values refer to the values of ($\textit{w}_0, \textit{w}_{a}$). The dashed vertical lines show the median convergence radius for haloes in that mass bin within which the density profiles should not be trusted. The error bars show the standard error on the median for the $\Lambda$CDM cosmology.}
    \label{fig:density_dm}
\end{figure*}

Fig.~\ref{fig:clustercounts} show the number counts for haloes with M$_{200,\rm crit} \geq$ 10$^{12}$M$_\odot$, 10$^{13}$M$_\odot$ and 10$^{14}$M$_\odot$ out to $z=3$ for the collisionless simulations for the different cosmologies. As expected from the HMF in Fig.~\ref{fig:hmf}, the number counts decrease for the freezing dark energy model and increase for the thawing dark energy models with increasing redshift relative to the $\Lambda$CDM cosmology. The crossing of the ratios in Fig.~\ref{fig:hmf} can also be seen in the ratios of number counts where haloes with M$_{200,\rm crit} \geq$ 10$^{13}$M$_\odot$ cross over at $z=1$. Because of the steepness of the HMF, the cluster count signal is dominated by the lowest-mass haloes, those near the lower mass limits in each mass bin. The bottom panels show that the signal is strongest for the highest-mass haloes and at higher redshifts. We plot error bars to show the Poisson uncertainties from the number of haloes in each redshift bin for the $\Lambda$CDM cosmology only for clarity, but note that the uncertainties for the other cosmologies are approximately of the same level. The uncertainties increase with increasing redshift and increasing mass since there are fewer haloes in those bins.

As discussed in Section~\ref{sec:cosmoparam} (see Fig.~\ref{fig:alens}), a tension exists between the constraints in the $\sigma_8-\Omega_{\rm m}$ parameter space from CMB data and various LSS statistics, including number counts. LSS generally prefers lower values of $S_8$, which results in fewer collapsed structures, compared to the value obtained from CMB data. As all of our cosmologies are consistent with CMB data by construction, any cosmology that suppresses the growth of structure relative to the $\Lambda$CDM cosmology could help to alleviate this tension. Interestingly, we find that there is a non-monotonic behaviour in the variation in $S_8$ of our cosmologies and the impact on number counts relative to $\Lambda$CDM.  For example, the freezing cosmology suppresses the abundance of the most massive clusters (Fig.~\ref{fig:massfraction} displays this most clearly) at a level that is comparable with that of the most extreme thawing models and yet the freezing model has a lower value of $S_8$ than the reference $\Lambda$CDM model while the most extreme thawing models have a larger value.  The mapping between $S_8$ and cluster abundance is therefore more complex for (CMB-constrained) DDE models than for $\Lambda$CDM.  Weak lensing, on the other hand, should provide a more direct constraint on $S_8$ than cluster abundances, as it measures the (projected) matter power spectrum.  Thus, in principle, the combination of cluster abundances and cosmic shear should be helpful in constraining the parameters of DDE.

\section{Halo structure}
\label{sec:halostructure}
Having investigated the overall abundance of haloes, we next examine the effect of DDE on the internal structure of haloes. The statistics we focus on are the spherically-averaged density profiles for haloes in a given mass range, the halo concentration-mass relation.

\subsection{Total mass density profile}
\label{sec:densityprofiles}

We calculate the median radial total mass density profiles in 15 logarithmically spaced radial bins in the range $0.01<\rm r/\rm \rm R_{200,\rm crit}\leq1$ and for haloes in mass bins of 0.5 dex width in the range M$_{200,\rm crit}$ = 10$^{13}$-10$^{15}$M$_\odot$. The densities are scaled by $r^2$ to reduce the dynamic range. 

Since the masses of haloes are affected by the DDE cosmologies, different populations of haloes are selected in each mass bin for different cosmologies. This makes any comparison of the direct effect of different DDE cosmologies on the structure of haloes convoluted. In order to compare like-for-like haloes, we match haloes across simulations to the $\Lambda$CDM cosmology (see Section \ref{sec:hmf}). Therefore, the mass bins correspond to M$_{200,\rm crit}$ from the matched haloes in the dark matter-only $\Lambda$CDM simulations. Equally, the R$_{200,\rm crit}$ values used to normalise the radial density profiles are those of the haloes that have been matched to, i.e. the R$_{200,\rm crit}$ values from the dark matter-only reference $\Lambda$CDM simulations.

Fig.~\ref{fig:density_dm} shows the median radial total mass density profiles for the collisionless simulations for the different DDE cosmologies for different mass bins and the ratios relative to $\Lambda$CDM for each mass bin. The vertical dashed lines show the median convergence radius for haloes in that mass bin for the $\Lambda$CDM cosmology. The convergence radius was calculated using the method described in \citet{power2003} (Equation 20) but with a convergence criterion of 0.177 as advocated by \citet{ludlow2019}. The effect of the freezing DDE cosmology is to increase the density of dark matter haloes whereas the thawing DDE cosmologies has the opposite effect, decreasing the density. The DDE cosmologies change the density by at most $\sim$10\% and the difference decreases in amplitude with increase in halo mass. There is also a radial dependence that shows an increase in density with increasing radius. We show the standard error on the median as error bars for the $\Lambda$CDM cosmology only, but note that the errors on the other cosmologies are approximately the same as for the $\Lambda$CDM case. 

These general trends in the density profiles can most likely be attributed to the difference in mass of the matched haloes. For example, haloes that show a higher density relative to their matched $\Lambda$CDM counterpart in Fig.~\ref{fig:density_dm} also show a relative increase in their M$_{200,\rm crit}$ in Fig.~\ref{fig:massfraction}. This is similar with what was found by M17 and S20. M17 shows that cosmologies with massive neutrinos lower the masses of dark matter haloes relative to their matched $\Lambda$CDM haloes and these cosmologies also show an almost radially independent suppression of the density profiles. In S20, cosmologies with running of the spectral index suppress (increase) mass growth for low-mass (high-mass) haloes and this was also reflected in their respective halo densities. 

\subsection{Concentration--mass relation}
\label{sec:massconcentration}

The internal structure of haloes are themselves tracers of the formation history of haloes. Since the formation history depends on the evolution of the background density, the internal structure is also sensitive to the cosmology. CDM models predict that low-mass haloes collapse earlier while high-mass haloes, such as clusters, are still collapsing today. As gravitational collapse can only occur when the local density exceeds the background density, lower-mass haloes are expected to have a more concentrated density profile, which has been shown to be true in a number of high resolution simulations (e.g. \citealt{bullock2001, eke2001}). 

Simulation results have shown that dark matter density profiles can be approximately described by the NFW profile \citep{navarro1997}

\begin{equation}\centering
    \rho(r) = \frac{\rho_{\rm s}}{\frac{r}{r_{\rm s}}[1+\frac{r}{r_{\rm s}}]^2}, 
\end{equation}

\noindent which only has a scale density, $\rho_{\rm s}$, and a scale radius, $r_{\rm s}$, as free parameters.  With this, one can define the concentration as c$_{200,\rm crit}\equiv \rm R_{200,\rm crit} / \rm r_{\rm s}$. 

To calculate the scale radius for our halo sample, we first remove all unrelaxed haloes as unrelaxed haloes are not in virial equilibrium and have been shown to be poorly described by the NFW profile (e.g. \citealt{maccio2007,romanodiaz2007}). This is done by ensuring that all haloes have their centre of mass offset by no more than 0.07 R$_{200,\rm crit}$ from the center of potential \citep{neto2007}, which has been shown to remove the vast majority of unrelaxed haloes \citep{duffy2008}. We select haloes with more than 800 particles and stack them until they have a total of 5000 particles. Lastly, we fit an NFW profile over the radial range 0.1 $\leq$ r/R$_{200,\rm crit}$ $\leq$ 1.0 and remove any halo for which $r_{\rm s} <$ convergence radius (described in Section~\ref{sec:densityprofiles}) to ensure the halo density profiles are converged. We fit to the quantity $\rho r^2$ to give equal weighting to each radial bin \citep{neto2007}.

\begin{figure}
	\includegraphics[width=\columnwidth]{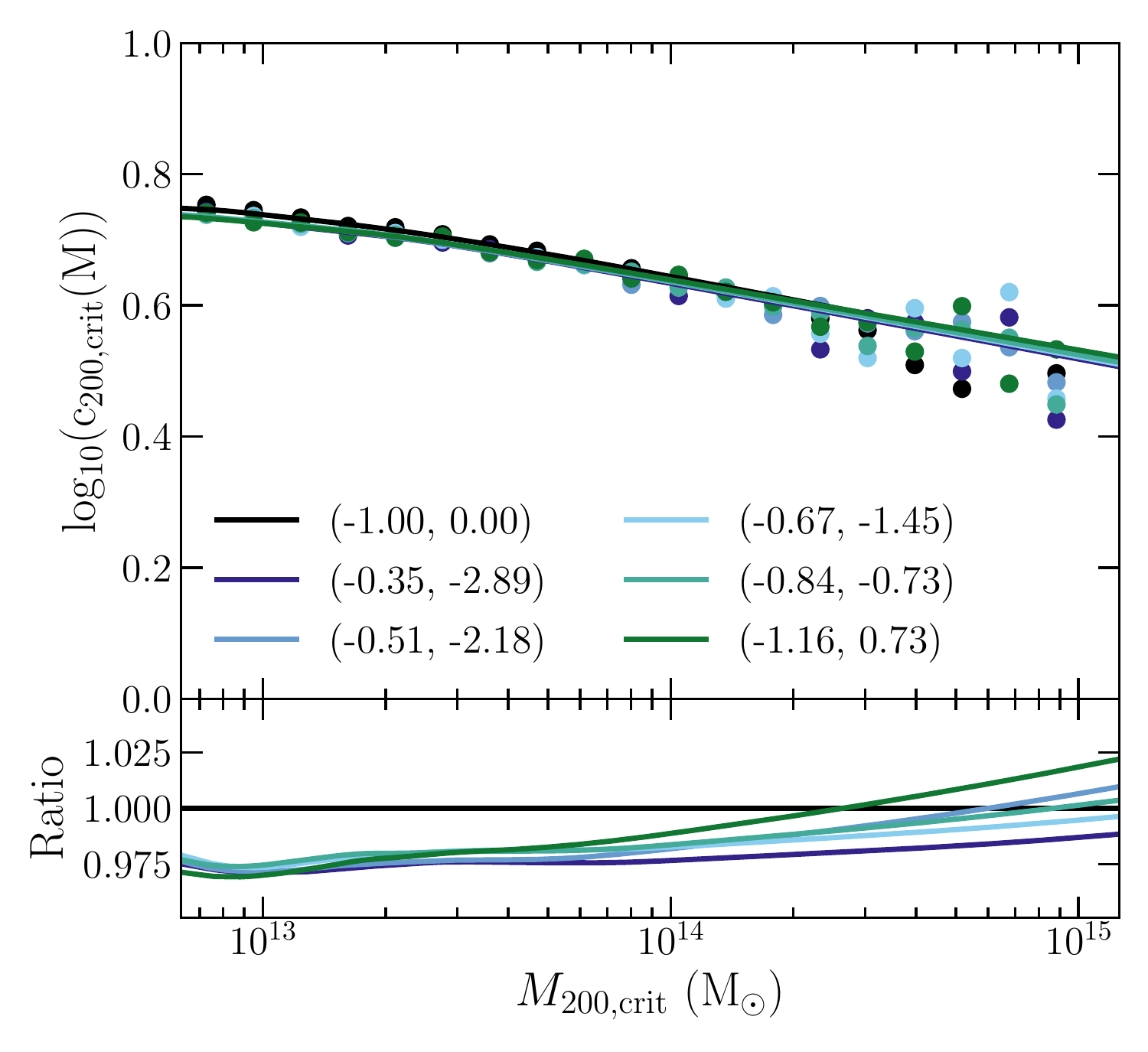}
    \caption{\textit{Top}: The concentration of dark matter haloes for the different cosmologies. Points represent the median concentration in 20 equally spaced bins between $10^{13} \leq$ M$_{200,\rm crit} < 10^{15.5}$ in bins of M$_{200,\rm crit}$ whereas the lines represent the LOWESS method applied to the unbinned data. \textit{Bottom}: The ratios of concentration of the LOWESS lines with respect to the $\Lambda$CDM cosmology. Colours indicate different cosmologies where bracketed values refer to the values of ($\textit{w}_0, \textit{w}_{a}$).}
    \label{fig:concentration_dm}
\end{figure}

In Fig.~\ref{fig:concentration_dm} we show the logarithm of the concentration for unmatched dark matter haloes for each cosmology (top). As we have not matched haloes, we use the M$_{200,\rm crit}$ and R$_{200,\rm crit}$ from each simulation. The dots represent the median value in each mass bin for 20 equally spaced bins between $10^{13} \leq$ M$_{200,\rm crit} < 10^{15.5}$ whereas the lines show the locally-weighted scatter plot smoothing (LOWESS) method \citep{lowess} applied to the unbinned data. The ratios were taken with respect to the $\Lambda$CDM cosmology and we use the lines from the LOWESS method rather than the binned median values (bottom). There is no strong trend in the ratios of the concentrations for the different cosmologies. Small differences appear at the high-mass end which is also where the scatter in the concentrations becomes significant.

M17 showed that massive neutrinos systematically lower the concentration of dark matter haloes at the $\approx$5-10\% level between the lowest and highest neutrino mass, while S20 showed that cosmologies with running of the spectral index increase the concentration towards higher masses for all considered cosmologies. M17 also showed that the effect of baryonic feedback dominates any effect on the concentrations which are affected by $\approx$20\%.

\section{Impact of baryons and its dependence on cosmology}
\label{sec:separability}
In this section we investigate the effects of including baryons on the statistics we have shown so far and show to what degree these can be separated from the effects of changing the cosmology. We use the term "separability" to refer to the degree by which cosmological and baryonic effects are independent of each other, or in other words, how sensitive one is to the other.

Many studies have shown that the inclusion of baryonic physics in cosmological simulation can have a significant effect on the overall matter distribution. This has been shown with respect to the matter power spectrum (e.g. \citealt{vandaalen2011,schneider2019,vandaalen2020}), the halo mass function (e.g. \citealt{sawala2013,cusworth2014,velliscig2014}), clustering \citep{vandaalen2014}, density profiles (e.g. \citealt{duffy2010,schaller2015}) and the binding energy of haloes \citep{davies2019}, which are all significantly impacted by baryons and their respective feedback mechanisms.  Of course, changes in cosmology also have a large effect on some of these statistics.  This raises the question of whether the effects of cosmology and baryons influence each other or, instead, can be treated independently, as often implicitly assumed in halo model-based approaches (e.g., \citealt{mead2016}). 

Such considerations are also important when constructing hydrodynamical simulations, since it is often desirable that they reproduce a particular set of observables. If those observables are sensitive to cosmological variations, then this would suggest that the simulations would need to be re-calibrated for each choice of cosmology.  The \texttt{BAHAMAS} suite of simulations are a first attempt at calibrating the feedback processes to study their impact on LSS for large-volume cosmological hydrodynamical simulations. It is therefore vital for the calibration statistics to be mostly unaffected by a change in cosmology, or to re-calibrate after every change. The calibration statistics for \texttt{BAHAMAS} are the observed stellar and hot gas mass of haloes, which were specifically chosen because they are expected to be relatively insensitive to changes in cosmology (as confirmed in \citealt{mccarthy2018}, S20, and later here) and because these quantities are directly related to impact of baryons on the matter power spectrum \citep{vandaalen2020,debackere2020}.  In the present study, we have used the same feedback parameters as adopted in \citet{mccarthy2017} and we have verified that the cosmologies considered in this work all reproduce the calibration statistics as well as found in that study.  

\begin{figure}
    \centering
    \includegraphics[width=\columnwidth]{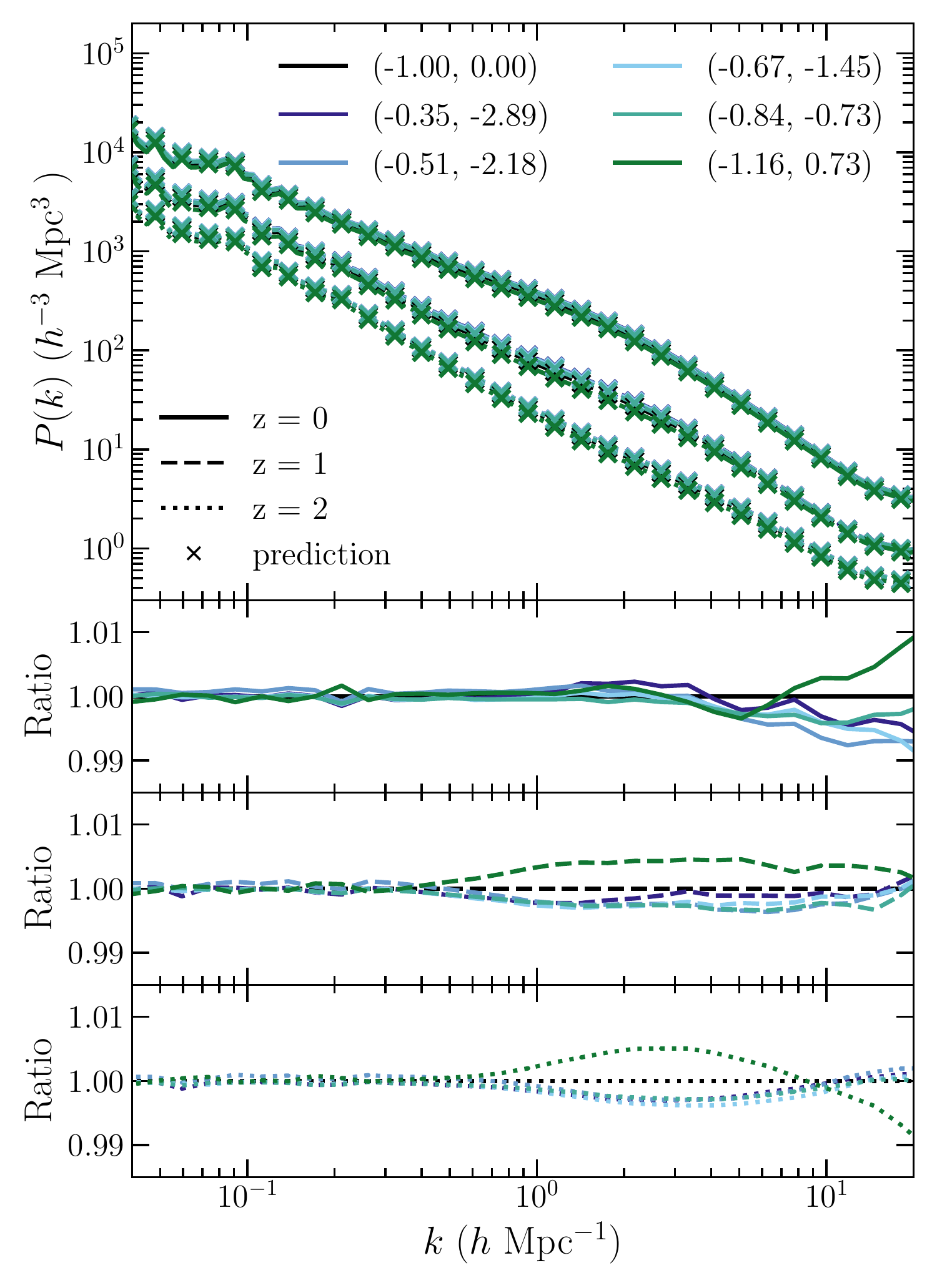}
    \caption{\textit{Top}: the total matter power spectra for the different cosmologies from hydrodynamical simulations (lines) and the collisionless simulations with added baryonic effects (crosses) as described in equation \ref{eq:separability}. Line styles indicate different redshifts. \textit{Bottom}: The ratios at different redshifts of the matter power spectrum from hydrodynamical simulations and the collisionless simulation with added baryonic effects for the same cosmology. Colours indicate different cosmologies where bracketed values refer to the values of ($\textit{w}_0, \textit{w}_{a}$) and linestyles indicate redshifts.}
    \label{fig:matterpowseparability}
\end{figure}

We have also confirmed that the relative impact of feedback (at fixed cosmology) on various metrics, such as the matter power spectrum and the halo mass function, are the same (to within a couple of percent) as reported previously in M17 and S20 (see also Fig.~\ref{fig:massfractionseparability} below).  Therefore, rather than re-examine the effects of baryons, we limit our exploration here to the question of whether the impact of baryons is separable from the change in cosmology.  To do so, we follow the approach taken in M17 and S20.  Specifically, if the effects are separable, then multiplying the impact of baryons in the reference $\Lambda$CDM run (relative to the collisionless version of this simulation) by the impact of changing the nature of dark energy relative to $\Lambda$CDM for the collisionless case, should reproduce the combined impact of baryons and a change in cosmology of a DDE run with hydrodynamics compared to the collisionless $\Lambda$CDM run.

To express the above mathematically, we test the {\it ansatz} that
\begin{equation}\label{eq:separability}
    \psi_{\rm H}^{\rm cosmo} = \psi_{\rm DMO}^{\rm \Lambda CDM} \left(\frac{\psi_{\rm DMO}^{\rm cosmo}}{\psi_{\rm DMO}^{\rm \Lambda CDM}}\right) \left(\frac{\psi_{\rm H}^{\rm \Lambda CDM}}{\psi_{\rm DMO}^{\rm \Lambda CDM}}\right).
\end{equation}

\noindent where $\psi$ is the chosen statistic (such as the matter power spectrum or the HMF), the subscripts denote whether it is from the collisionless or hydrodynamical cases, and the superscripts denote the cosmology where `cosmo' refers to either a DDE or $\Lambda$CDM cosmology. The first and second bracketed terms are therefore the effect of cosmology and baryons with respect to a collisionless (dark matter-only) $\Lambda$CDM cosmology simulation, respectively. 

To test the separability, we have run all of the simulations including hydrodynamics, using the calibrated feedback model from the original \texttt{BAHAMAS} runs \citep{mccarthy2017}.  We test the degree of separability by applying equation \ref{eq:separability} to three statistics: the matter power spectrum, the HMF and the density profiles. 

\begin{figure}
    \centering
    \includegraphics[width=\columnwidth]{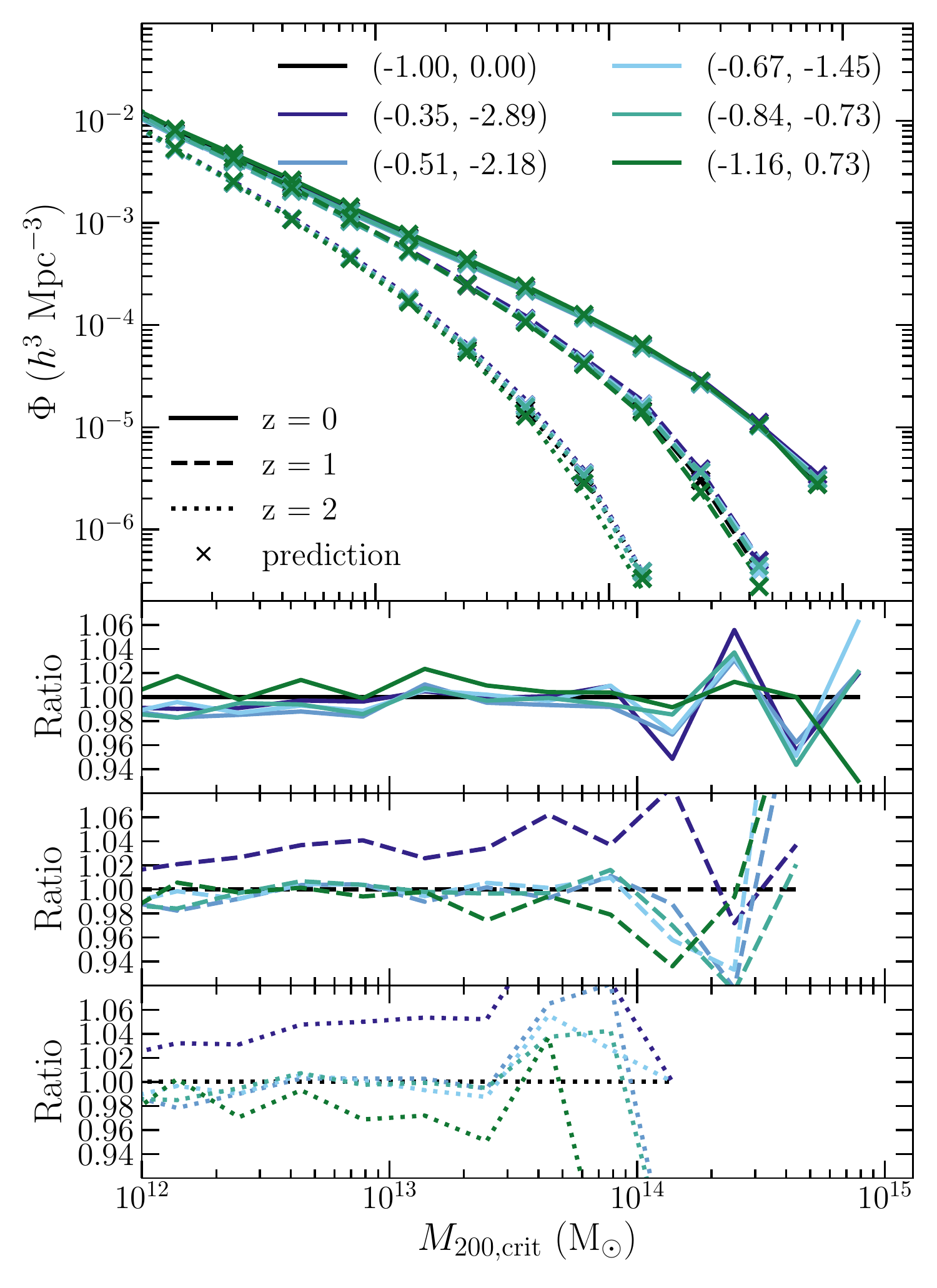}
    \caption{\textit{Top}: the HMF for the hydrodynamical simulations (lines) and the collisionless simulations with added baryonic effects (crosses) as described in Equation~\ref{eq:separability}. Line styles indicate different redshifts. \textit{Bottom}: The ratios at different redshifts of the HMF from hydrodynamical simulations and the collisionless simulation with added baryonic effects for the same cosmology. Colours indicate different cosmologies where bracketed values refer to the values of ($\textit{w}_0, \textit{w}_{a}$) and linstyles indicate redshifts.}
    \label{fig:hmfseparability}
\end{figure}

\subsection{Matter power spectrum}
We first examine the total matter power spectrum, which was described in Section \ref{sec:matterclustering}. 
Fig.~\ref{fig:matterpowseparability} shows the total matter power spectrum from the hydrodynamical simulations (lines) and from the collisionless simulations with baryonic effects applied following the prescription in equation \ref{eq:separability} (crosses). To see how well these agree, we show the ratio for each cosmology and at three different redshifts, $z=0$, $1$, and $2$.  As can be seen, the effects of cosmology and baryons (feedback) are separable to very high precision (typically $<0.1\%$ for $k<10$ $h$ Mpc$^{-1}$) for the majority of the $k$-scales and across all redshifts  shown.

\subsection{Halo mass function}
Next we examine the separability of baryonic and cosmological effects on the HMF, which was described in Section \ref{sec:hmf}. Fig.~\ref{fig:hmfseparability} shows the HMF for the hydrodynamical simulation (lines) and the collisionless simulations including baryonic effects following the prescription in equation \ref{eq:separability} (crosses). As with the matter power spectrum, we show the ratio of these for each cosmology and for $z=0$, $1$, $2$. They typically agree to better than a few percent accuracy for the majority of the mass range sampled at each redshift.  The scatter increases somewhat at the high-mass end at each redshift due to the relative rarity of such systems.

\begin{figure}
    \centering
    \includegraphics[width=\columnwidth]{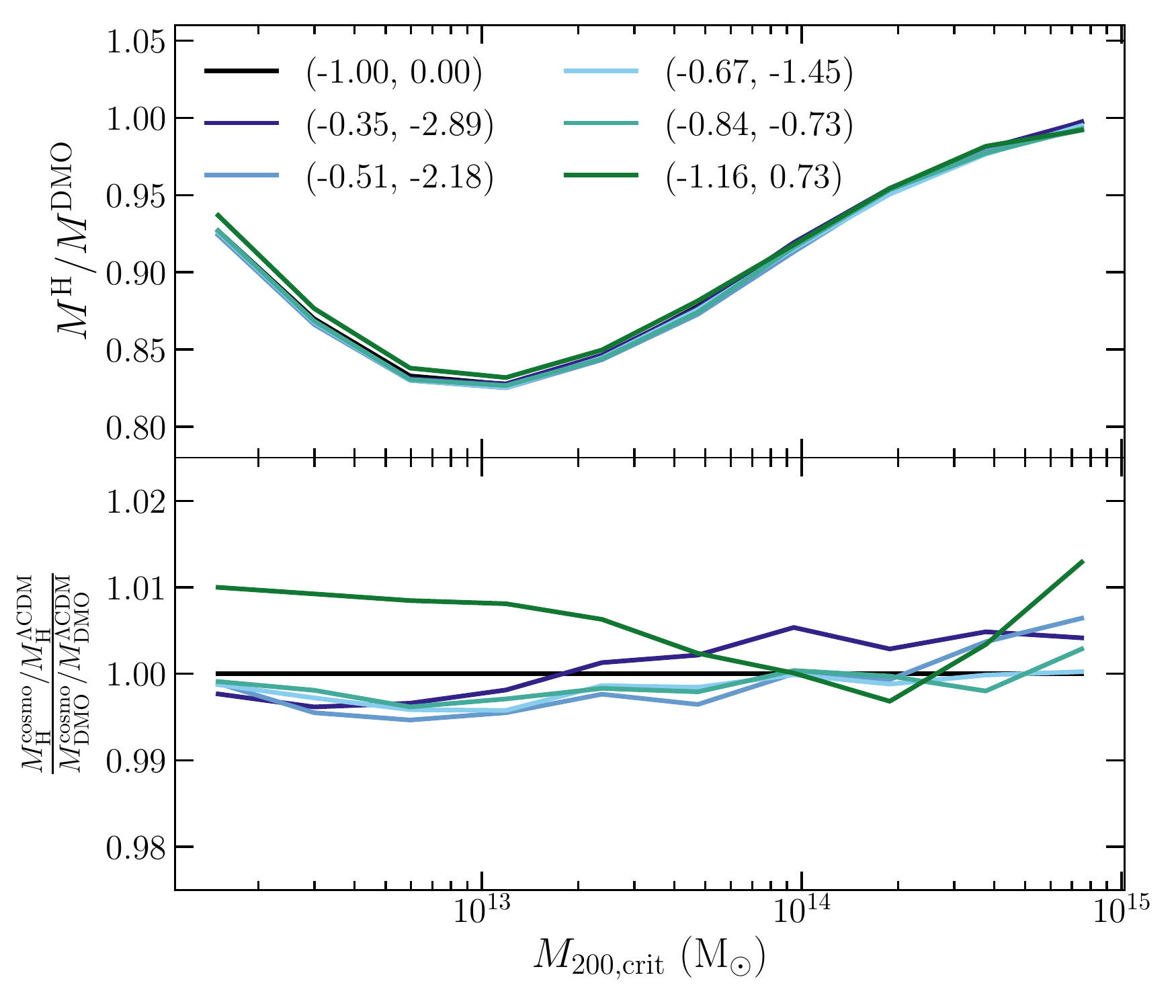}
    \caption{\textit{Top}: The fractional change in halo mass, M$_{200,\rm crit}$, of haloes from the hydrodynamical simulations relative to their matched dark matter counterparts at $z=0$. \textit{Bottom}: The ratio of the fractional mass change from hydrodynamical simulations and the collisionless simulations with added baryonic effect (see Equation~\ref{eq:separability}). Colours indicate different cosmologies where bracketed values refer to the values of ($\textit{w}_0, \textit{w}_{a}$).}
    \label{fig:massfractionseparability}
\end{figure}

We also investigate the separability of cosmological and baryonic effects on the masses of haloes, as we have shown that the halo mass is affected by DDE (Fig.~\ref{fig:massfraction}) and previous studies have shown the impact of baryonic physics on halo mass \citep{sawala2013,cui2014,velliscig2014,schaller2015}.  The top panel of Fig.~\ref{fig:massfractionseparability} shows the ratio of M$_{200,\rm crit}$ from the hydrodynamical simulations with those of the collisionless simulations in bins of M$_{200,\rm crit}$ of the halo from the collisionless simulation, for each cosmology. It shows that baryonic effects suppress the masses of haloes by up to $\approx$15$\%$ at 10$^{13}$M$_\odot$ but this suppression is less effective at lower and higher masses, consistent with M17 and S20. This peak in suppression is due to the mass dependence of the feedback efficiency of active galactic nuclei (AGN).  The suppression is reduced in magnitude at higher masses owing to the increased binding energies of those haloes. In the bottom panel of Fig.~\ref{fig:massfractionseparability} we plot the effect of the DDE cosmologies on the halo mass for the hydrodynamical simulations normalised by the effect of DDE in the collisionless simulations (for each cosmology).  The impact of baryons on the halo is independent of the nature of DDE at the level of $<1\%$ over the entire mass range.  Likewise, the effect of DDE cosmologies on halo mass is independent of baryonic physics.

\subsection{Total matter density profiles}

\begin{figure*}
    \includegraphics[width=\textwidth]{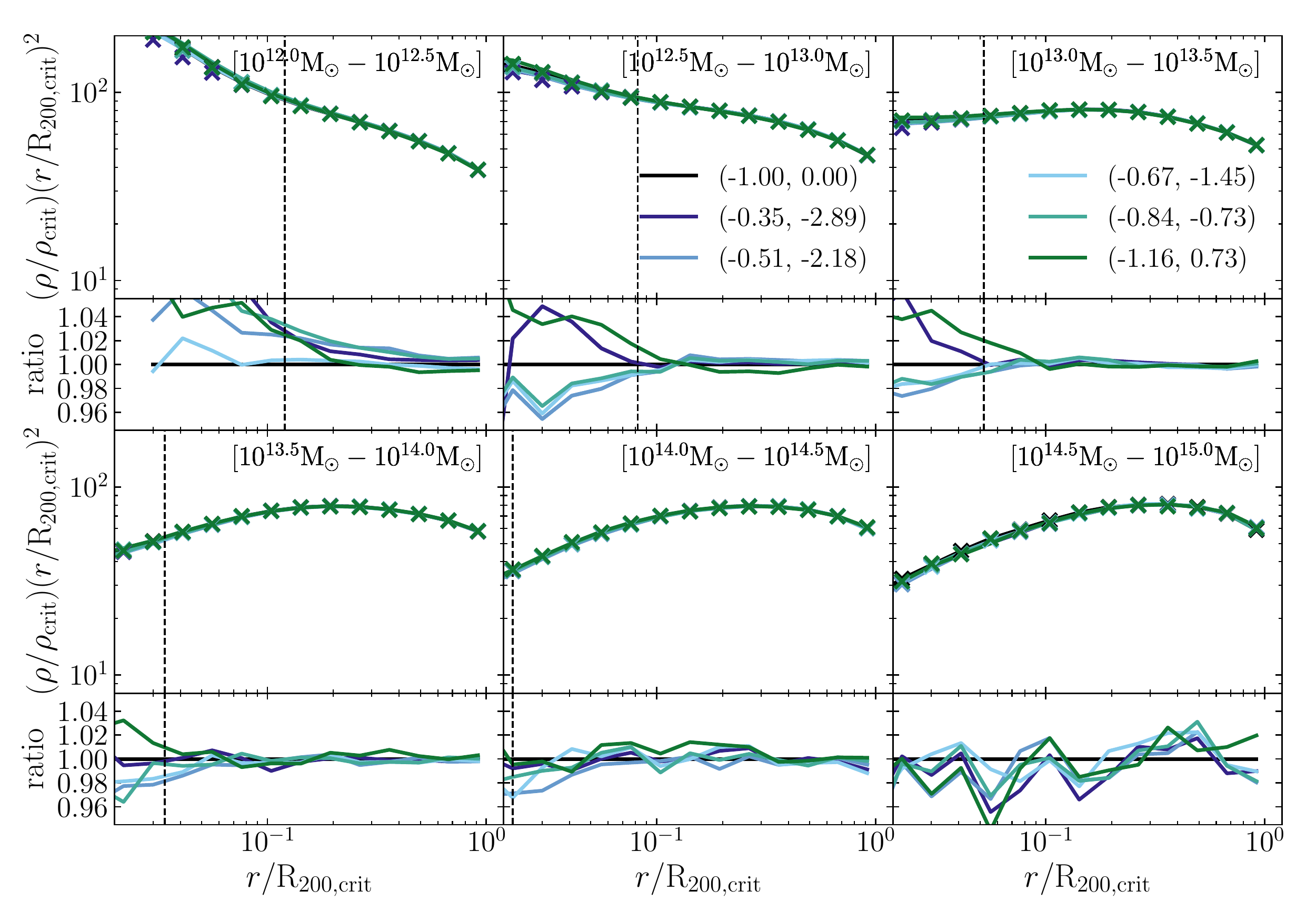}
    \caption{Median radial total mass density profiles for the different DDE cosmologies from the hydrodynamical simulations (lines) and the collisionless simulations with added baryonic effects (crosses) as described by Equation~\ref{eq:separability}. The ratio is of the hydrodynamical density profiles and the collisionless density profiles with added baryonic effects. Each panel shows different mass bins with width 0.5 dex in M$_{200,{\rm crit}}$. Colours indicate different cosmologies where bracketed values refer to the values of ($\textit{w}_0, \textit{w}_{a}$). The dashed vertical lines show the median convergence radius (see Section~\ref{sec:densityprofiles}) for haloes in that mass bin within which the density profiles should not be trusted.}
    \label{fig:densityseparability}
\end{figure*}

Finally, we examine the separability of cosmological and baryonic effects on the total matter density profiles, described in Section~\ref{sec:densityprofiles}. Fig.~\ref{fig:densityseparability} shows the total matter density profiles in bins of M$_{200,\rm crit}$ of haloes from the hydrodynamical simulation (lines) and of the collisionless simulations where the baryonic effects (crosses) are applied in post-processing according to Equation~\ref{eq:separability}. Unlike in Fig.~\ref{fig:density_dm}, these haloes have not been matched to the collisionless simulations. 

The effects of feedback and changes in cosmology are separable to $<1\%$ for haloes within the mass range M$_{200,\rm crit}$=$10^{12.5}-10^{14}$ M$_{\odot}$. The errors in the separability are slightly larger for lower-mass haloes (likely because they are sampled by fewer particles), and at the highest masses, plausibly as a result of relatively poor statistics.

\section{Discussion and Summary}
\label{sec:discussionresults}
We have constructed a new suite of cosmological hydrodynamical simulations using a modified version of the \texttt{BAHAMAS} code to investigate the effects of spatially flat DDE cosmologies on LSS.  Six cosmologies were chosen based on the constrained $\textit{w}_0 - \textit{w}_{a}$ geometric degeneracy from the \textit{Planck} TT+lowTEB data set. We included $A_{\rm lens}$ as a free parameter in our analysis to account for the enhanced smoothing of the CMB temperature power spectrum. DDE changes the expansion history of the Universe (see Fig.~\ref{fig:expansionhistory}) and therefore affects the growth of structure. However, we choose the other cosmological parameters so that the {\it integrated} expansion history (i.e., the distance to the surface of last scattering) is the same and consistent with the {\it Planck} primary CMB data. While this approach generates more `realistic' cosmologies, it makes disentangling the effects of DDE from those caused by changes in the other cosmological parameters more challenging.  Therefore, we refer to the DDE cosmologies as a whole, rather than the DDE itself, and all effects are with respect to our $\Lambda$CDM cosmology. While our analysis is restricted to a single extension to $\Lambda$CDM, as we want to see the effects of DDE, we do compare the results to the possible cosmological extensions of a running of the spectral index as well as changes to the summed neutrino mass. It would be interesting to  let DDE, running, and massive neutrinos vary simultaneously (e.g., see \citealt{divalentino2016,divalentino2020}) to examine the degeneracies between their effects, but we leave this for future studies. To examine the impact of the DDE cosmologies on the LSS, we have examined a variety of statistics, namely: the matter power spectrum, the 2-point auto-correlation function of dark matter haloes, the halo mass function, and halo number counts.  We also examined the density profiles and the concentration--mass relation to investigate the impact of DDE on internal properties of haloes.

Our main findings can be summarised as follows:
\begin{itemize}

    \item  The clustering of matter is strongly affected in the DDE cosmologies.  Both the matter power spectrum (Fig.~\ref{fig:matterpowerspec}) and the 2-point auto-correlation function of haloes (Fig.~\ref{fig:2pcf}) can show up to a $\sim$10\% change at $z=0$, where the thawing (freezing) DDE cosmologies enhance (suppress) the clustering with respect to the reference $\Lambda$CDM cosmology. The effect on $P(k)$ shows only a weak scale dependence, while the amplitude change agrees well with the expectations based on changes in the linear growth factor for the different cosmologies.  The redshift dependence of these effects is relatively mild.
    
    \item The effect on the abundance of low-mass haloes in the different DDE cosmologies is of the same order of magnitude as the clustering and has a strong mass dependence (Fig.~\ref{fig:hmf}). The largest effects are seen at the high-mass end and at higher redshifts, $z=1$ and $z=2$. The abundances of the lowest-mass haloes in our simulations at any given redshift are modified by $\approx$5-10\% while the highest-mass haloes can have their abundances modified by up to $\approx$20\%. The thawing (freezing) DDE cosmologies decrease (increase) the abundance of low-mass haloes with respect to $\Lambda$CDM, whereas for high-mass haloes (M$_{200,{\rm crit}} \ga 10^{14}$ M$_\odot$) this trend is reversed. The effect at the low-mass end can be attributed to the differences in $\Omega_{\rm m}$ between the cosmologies, while the changes at the high-mass end are due to the change in growth of structure (i.e., $P(k)$).
    
    \item In terms of the internal structure of haloes, the DDE cosmologies generally have less of an impact. The density profiles show shifts in amplitude consistent with the change in halo mass mentioned above (Fig.~\ref{fig:density_dm}), while the shapes of the density profiles are only weakly affected. 

    \item The effects on our chosen statistics have been compared to the effect of massive neutrinos with a varying sum of neutrino masses previously investigated in BAHAMAS \citep{mummery2017}. It is clear that massive neutrinos can behave similarly to the cosmologies including DDE presented here. However, massive neutrinos can only suppress the clustering of matter and haloes and the abundances of haloes, whereas the freezing and thawing cosmologies can either suppress or enhance these, respectively. Their scale and redshift dependence on $P(k)$, as well as their mass and redshift dependence on the abundance of haloes, are very similar which makes the effect of massive neutrinos and thawing cosmologies difficult to distinguish (see also \citealt{upadhye2019}).
    
    \item We have also compared these statistics to the effect of a running of the scalar spectral index found by \citep{stafford2020}. The comparison to running of the spectral index shows striking similarities in both the shape and magnitude of observed trends for all the considered statistics. The scale and redshift dependencies in $P(k)$ are very similar for the scales we sample although there appears to be some deviations in behaviour at the smallest scales, largest $k$. The effect on the abundance of dark matter haloes is even more similar, with both cosmological extensions showing remarkably similar trends across mass and redshift. 
    
    \item These effects of changing cosmology were also compared to the effects of baryons. In general, baryons tend to suppress the statistics considered here, at levels of up to 10-20\%. Baryons also have a strong scale dependence for $P(k)$ and a more complicated mass dependence for the HMF compared to that of DDE cosmologies. We investigated the separability of cosmological and baryonic effects on our LSS statistics, by assuming that each effect can be treated as a simple multiplicative factor described by Equation~\ref{eq:separability}.  In general, we find that effects due to the different DDE cosmologies and baryonic physics can be separated to high accuracy in this way, with errors of at most a few percent.  More specifically, errors in the separability in $P(k)$ are $<0.1\%$ (see Fig.~\ref{fig:matterpowseparability}), the lowest in any of our statistics, while in the HMF and density profiles the errors are typically $\approx$1-2\% (see Fig.~\ref{fig:hmfseparability}).

\end{itemize}{}

We can put our work in the context of the LSS tension that exists between `early-Universe' CMB data and `late-Universe' LSS data sets, where the latter prefer lower values of $S_8$ than the former, which is effectively equivalent to saying that the observed low-redshift Universe is smoother than it ought to be assuming the $\Lambda$CDM model with CMB constraints on its parameters.  It is clear from Table~\ref{tab:cosmoparams} that our cosmologies do not significantly lower the $S_8$ parameter relative to $\Lambda$CDM and most DDE models we consider actually increase its value (the thawing models).  This simple comparison, though, relies on the assumption that changes to the value of $S_8$ directly translate into changes in the formation of structure. However, we have shown that for (CMB-constrained) DDE models, the mapping between massive cluster abundances in particular and $S_8$ is more complex (non-monotonic) than for $\Lambda$CDM, with the freezing model (with lower $S_8$) yielding a similar suppression in cluster abundances to some of the extreme thawing models (with higher $S_8$).  Cosmic shear (weak lensing), on the other hand, might be expected to more directly constrain $S_8$, given that it measures the projected matter power spectrum.  In principle, therefore, the combination of different LSS tests should be helpful in constraining the nature of DDE.  As for the current claimed tension between the CMB and measures of LSS, the variations we see in the DDE cosmologies, while certainly not insignificant, do not appear to be large enough on their own to reconcile the tension (e.g., the abundance of clusters is suppressed by $\approx5\%$ for some of the models, whereas a suppression of $\approx50\%$ or larger is claimed to be required, depending on how the mass scale of clusters is calibrated).

In conclusion, the impact of DDE in CMB-constrained cosmologies results in significant effects on a variety of LSS metrics which should be testable with upcoming LSS surveys.  For example, LSST\footnote{\url{https://www.lsst.org/}} and Euclid\footnote{\url{https://www.euclid-ec.org/}} are anticipated to measure the matter power spectrum at the percent level \citep{huterer2002,huterer2005,hearin2012}, while the differences in the DDE models we consider can reach up to ten times this level.  The prospects for using future LSS observations together with detailed predictions from cosmological simulations to place interesting constraints on DDE are therefore bright.

\section*{Acknowledgements}

SP acknowledges an STFC doctoral studentship. This project has received funding from the European Research Council (ERC) under the European Union's Horizon 2020 research and innovation programme (grant agreement No 769130).
This work used the DiRAC@Durham facility managed by the Institute for Computational Cosmology on behalf of the STFC DiRAC HPC Facility. The equipment was funded by BEIS capital funding via STFC capital grants ST/P002293/1, ST/R002371/1 and ST/S002502/1, Durham University and STFC operations grant ST/R000832/1. DiRAC is part of the National e-Infrastructure.

\section*{Data availability}
The data underlying this article will be shared on reasonable request to the corresponding author.


\bibliographystyle{mnras}
\bibliography{biblio}


\appendix

\section{}\label{sec:appendixa}

In this section we show how we can use the HMF fitting function from \citet{tinker2008} to gain some insight into the trends we observe in the HMF for our different cosmologies in Fig.~\ref{fig:hmf}. 

The fitting function provided in \citet{tinker2008} is an attempt to describe the abundance of dark matter haloes as a function of the matter power spectrum. It is given by

\begin{equation}\label{equ:tinker}
    \frac{dn}{dM}=f(\sigma)\frac{\bar\rho_{\rm m}}{M}\frac{d\rm ln \sigma^{-1}}{dM}
\end{equation}

\noindent where $\bar\rho_{\rm m}$ is the mean matter density which depends on $\Omega_{\rm m}$, and $f(\sigma)$ is given as 

\begin{equation}\label{equ:tinkerfit}
    f(\sigma)=A\left[\left(\frac{\sigma}{b}\right)^{-a}+1\right]e^{-c/\sigma^2}
\end{equation}

\noindent where $A$, $a$, $b$, and $c$ are constants calibrated to simulations, and $\sigma$ is the rms density fluctuation in a sphere of radius $R$,

\begin{equation}\label{equ:sigma}
    \sigma^2(R)=\frac{1}{2\pi^2}\int_{0}^{\infty}k^2P(k)|W(kR)|^2dk,
\end{equation}

\noindent where $W(kR)$ is the Fourier transform of the real-space top-hat window function and $P(k)$ is the linear matter power spectrum. 

\begin{figure}
    \centering
    \includegraphics[width=\columnwidth]{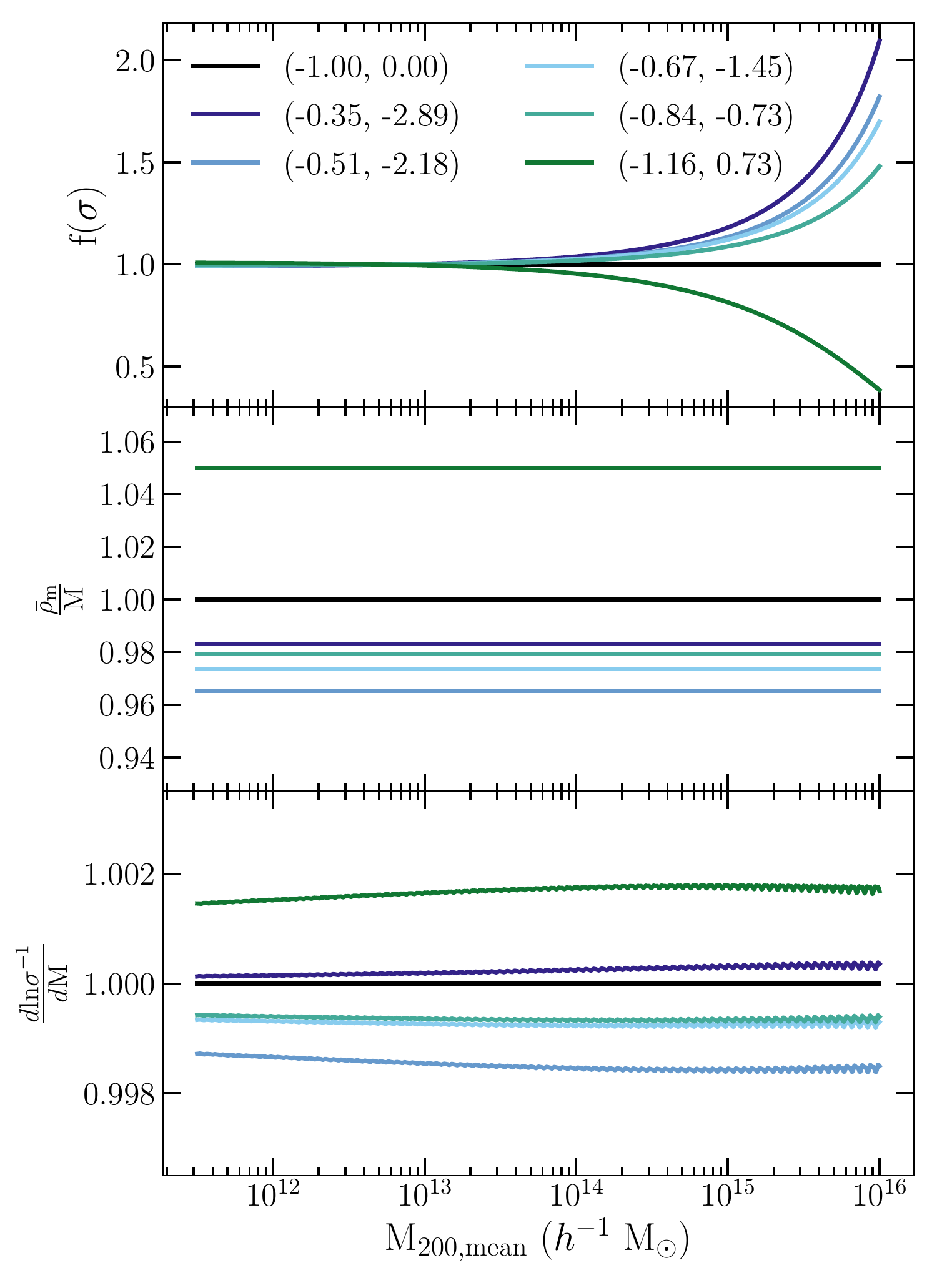}
    \caption{The HMF fitting function from \citet{tinker2008} given in Equation~\ref{equ:tinker} split into three separate terms and plotted against halo mass. Each line represents one of our cosmologies normalised by the $\Lambda$CDM cosmology at $z=0$. Colours indicate different cosmologies (see Table~\ref{tab:cosmoparams}) where bracketed values refer to the values of ($\textit{w}_0, \textit{w}_{a}$).}
    \label{fig:tinker}
\end{figure}

Through Equation~\ref{equ:tinker}, we can decompose the abundance of haloes into three separate terms; $f(\sigma)$ which depends on a set of constants and a cosmology-dependent $P(k)$, $\frac{\bar\rho_{\rm m}}{M}$ which is cosmology dependent through its $\Omega_{\rm m}$ dependence, and $\frac{d\rm ln \sigma^{-1}}{dM}$ which is also dependent on cosmology through $P(k)$. To investigate the impact that each of these terms has on the final HMF, we show them as a function of halo mass in Fig.~\ref{fig:tinker} for our cosmologies at $z=0$, normalised by their respective $\Lambda$CDM cosmology solution. The $f(\sigma)$ term changes the abundance of high-mass haloes that are still forming at $z=0$ but leaves the low-mass end unaffected. The $\frac{\bar\rho_{\rm m}}{M}$ term, which is effectively just a change in $\Omega_{\rm m}$, creates a constant offset equal to the fractional difference between the values of $\Omega_{\rm m}$ for the cosmologies compared to the value of the $\Lambda$CDM cosmology. The $\frac{d\rm ln \sigma^{-1}}{dM}$ term also shows an almost constant, but negligible, offset. Fig.~\ref{fig:tinker} can then be used to explain the trends we see in the HMF for the different cosmologies in Fig.~\ref{fig:hmf}. Firstly, the low-mass trend is dominated by changes in $\Omega_{\rm m}$ between the cosmologies and, secondly, the effect at the high-mass end is a combination of this offset and changes in $P(k)$ that cause a suppression or enhancement of the HMF at the high-mass end.

\begin{figure}
    \centering
    \includegraphics[width=\columnwidth]{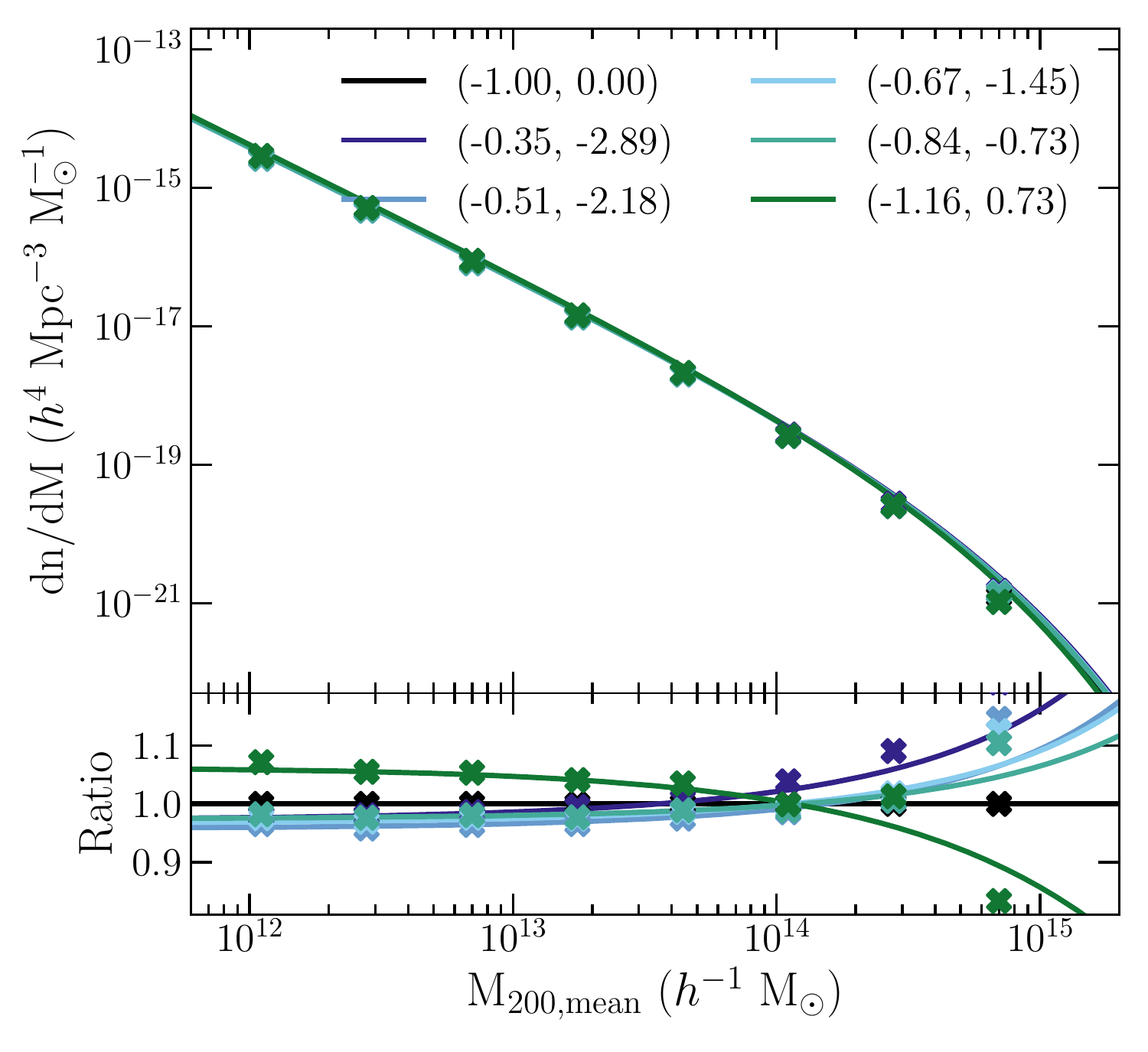}
    \caption{\textit{Top:} The HMF fitting function described in Equation~\ref{equ:tinker} (lines) and the HMF from the collisionless simulations (crosses) for each of the cosmologies at $z=0$. \textit{Bottom:} The ratios of the fitting function and the simulation HMF relative to their respective $\Lambda$CDM cosmology solution. Colours indicate different cosmologies (see Table\ref{tab:cosmoparams}) where bracketed values refer to the values of ($\textit{w}_0, \textit{w}_{a}$).}
    \label{fig:tinkersim}
\end{figure}

For completeness we also compare the HMF from the simulations to the results from the fitting function of \citet{tinker2008} in Fig.~\ref{fig:tinkersim}. The ratios have been taken with the $\Lambda$CDM cosmology for the collisionless simulations (crosses) and the HMF fitting function (lines), respectively. We see that the fitting function reproduces the relative difference between the cosmologies well and over the entire mass range probed by the simulations, although there is some scatter at the high-mass end for the simulation results which make the direct comparison more challenging. 

\bsp	
\label{lastpage}
\end{document}